\documentclass[intlimits,twoside,a4paper]{article}

\usepackage{amsmath,amssymb}
\usepackage{graphicx}
\usepackage{wrapfig}

\usepackage[T2A]{fontenc}
\usepackage[cp1251]{inputenc}

\usepackage{color}

\usepackage[eqsecnum]{cmpj2}

\usepackage{url}
\issue{2012}{15}{4}{40002}

\doinumber{10.5488/CMP.15.40002}

\title[Lviv period for Smoluchowski]{Lviv period for Smoluchowski: Science, teaching, \\ and beyond}

\author[A. Rovenchak]{A. Rovenchak}

\address{Department for Theoretical Physics,
Ivan Franko National University of Lviv, \\
12 Drahomanov St., 79005 Lviv, Ukraine
}

\authorcopyright{A. Rovenchak, 2012}
\date{Received October 30, 2012, in final form November 21, 2012}

\begin{document}

\maketitle

\begin{abstract}
A major part of Marian Smoluchowski's achievements in science corresponds to the period of his work at the University of Lviv. Since this part is well described in the literature, in the paper the emphasis is made on some less known activities of this outstanding scientist: his teaching, his organizational efforts, and even his hobbies. The list of publications corresponding to the Lviv period is given.

\keywords Marian Smoluchowski, University of Lviv

\pacs
01.30.Tt, 
01.60.+q, 
01.65.+g. 
\end{abstract}

\def\be{\begin{eqnarray}}
\def\ee{\end{eqnarray}}
\def\ben{\begin{eqnarray*}}
\def\een{\end{eqnarray*}}

\def\eps{\varepsilon}

\section{Short biography}

\begin{wrapfigure}{r}{0.33\textwidth}
\vspace*{-0.4cm}
\centerline{\includegraphics[clip,width=0.33\textwidth]{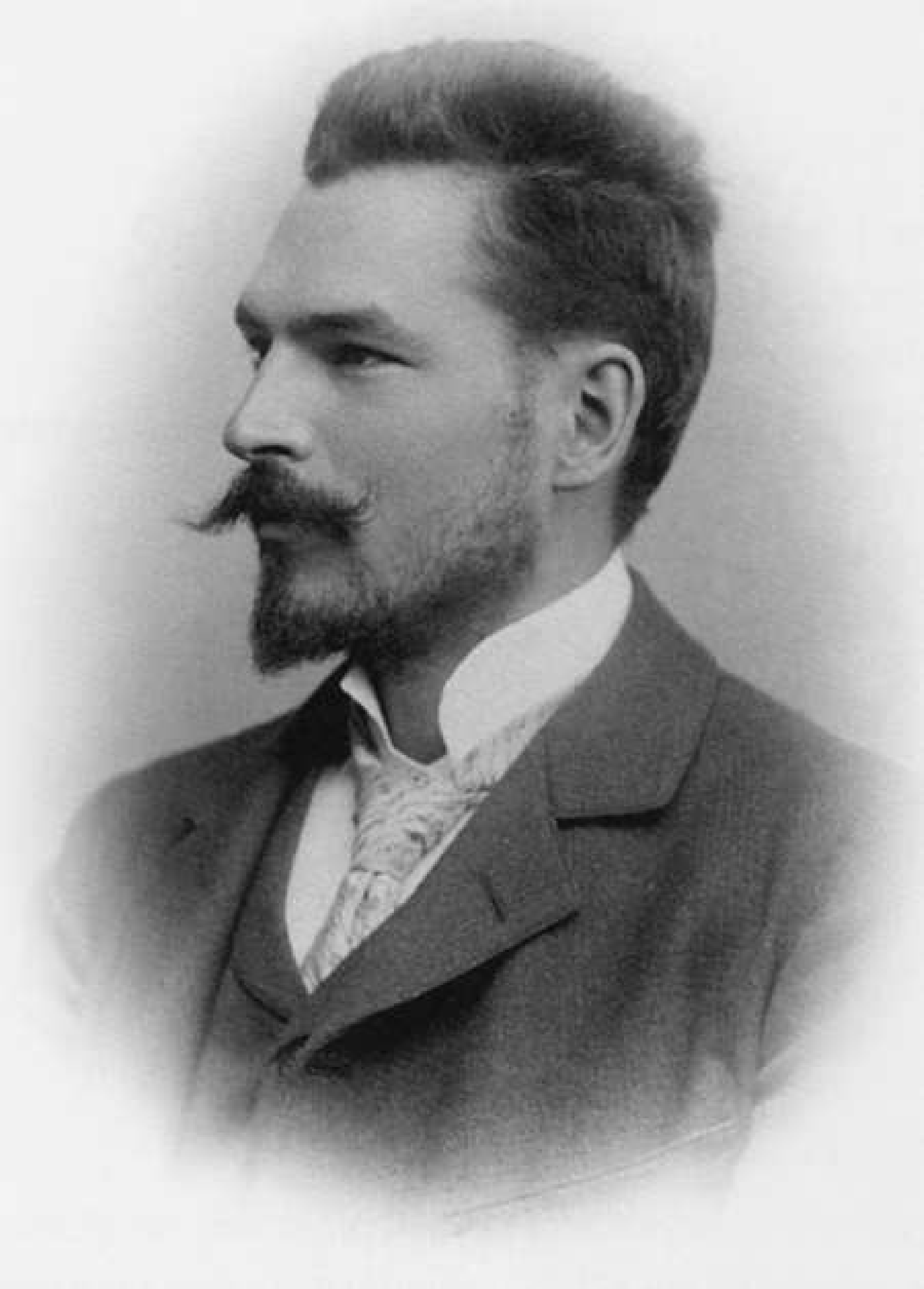}}
\caption{Marian Smoluchowski during Lviv period. Photo by E.~Trzemeski.}
\label{fig:portrait}
\end{wrapfigure}
Marian Smoluchowski (Maryan Ritter\footnote{Ritter is a title of nobility in German-speaking areas, in particular in the Austrian Empire. It is translated as ``knight'' and refers to the second lowest rank nearly equal to Baronet (below Baron).} von Smolan Smoluchowski) was born
on May, 28, 1872 in Vorder-Br\"uhl (M\"odling, near Vienna, now a suburb) to the family of Wilhelm Smoluchowski, a high official in the chancellery of emperor Franz Joseph I, and Teofila Szczepanowska.
During 1880--90 he studied in Collegium Theresianum and upon finishing it obtained certificate with distinction \cite{Teske55,Teske77,Teske68,Sredniawa85,Chandr00}.

He continued his education at the University of Vienna studying physics in 1891--94. After army service in 1894/95, Marian Smoluchowski obtained in 1895 a doctoral degree with highest honors (\textit{sub Summis Auspiciis Imperatoris}). The title of his doctoral thesis was ``Akustische Untersuchungen \"uber die Elasticit\"at weicher K\"orper'' (Acoustic studies of the elasticity of soft bodies; advisor: Jo\v{z}ef  Stefan) \cite{Pohl08}.

Afterwards, he spent several years at universities abroad:
1895/96 in Paris, 1896/97 in Glasgow, 1897 in Berlin. Upon returning to Vienna in 1897, Marian Smoluchowski received \textit{veniam legendi} (right to teach), habilitating as a ``Privatdozent'' in physics.

In 1899 Smoluchowski moved to the University of Lviv, where he spent almost fourteen~--- most productive in his career~--- years.

In 1913 he accepted the invitation to head the Department for Experimental Physics at the Jagiellonian University (Krak\'ow).
In 1916/17 Smoluchowski was Dean of the Philosophy Faculty, and in June, 1917 he was elected Rector of the Jagiellonian University.
Unfortunately, he never started these duties: on September, 05, 1917 Marian Smoluchowski passed away.

\section{Starting work in Lviv}
Marian Smoluchowski moved to the University of Lviv\footnote{%
The city, also known as \textit{Lw\'ow} in Polish and \textit{Lemberg} in German, belonged to the Austrian-Hungarian Empire at that time.}, being persuaded by Kazimierz Twardowski (1866--1938), a Polish philosopher and logician, who also attended Collegium Theresianum \cite{Hunger08}. Smoluchowski was thought to lecture subjects in theoretical and mathematical physics replacing Professor Oskar Fabian, the first head of the Department for Theoretical Physics, who was already seriously ill at that time and died later in October 1899 \cite{Rovenchak09}.

Smoluchowski's request to the University Collegium of Professors is reproduced below \cite{ArchiveLU}:

\begin{table}[h!]
\begin{tabular}{cl}
\includegraphics[width=0.43\textwidth]{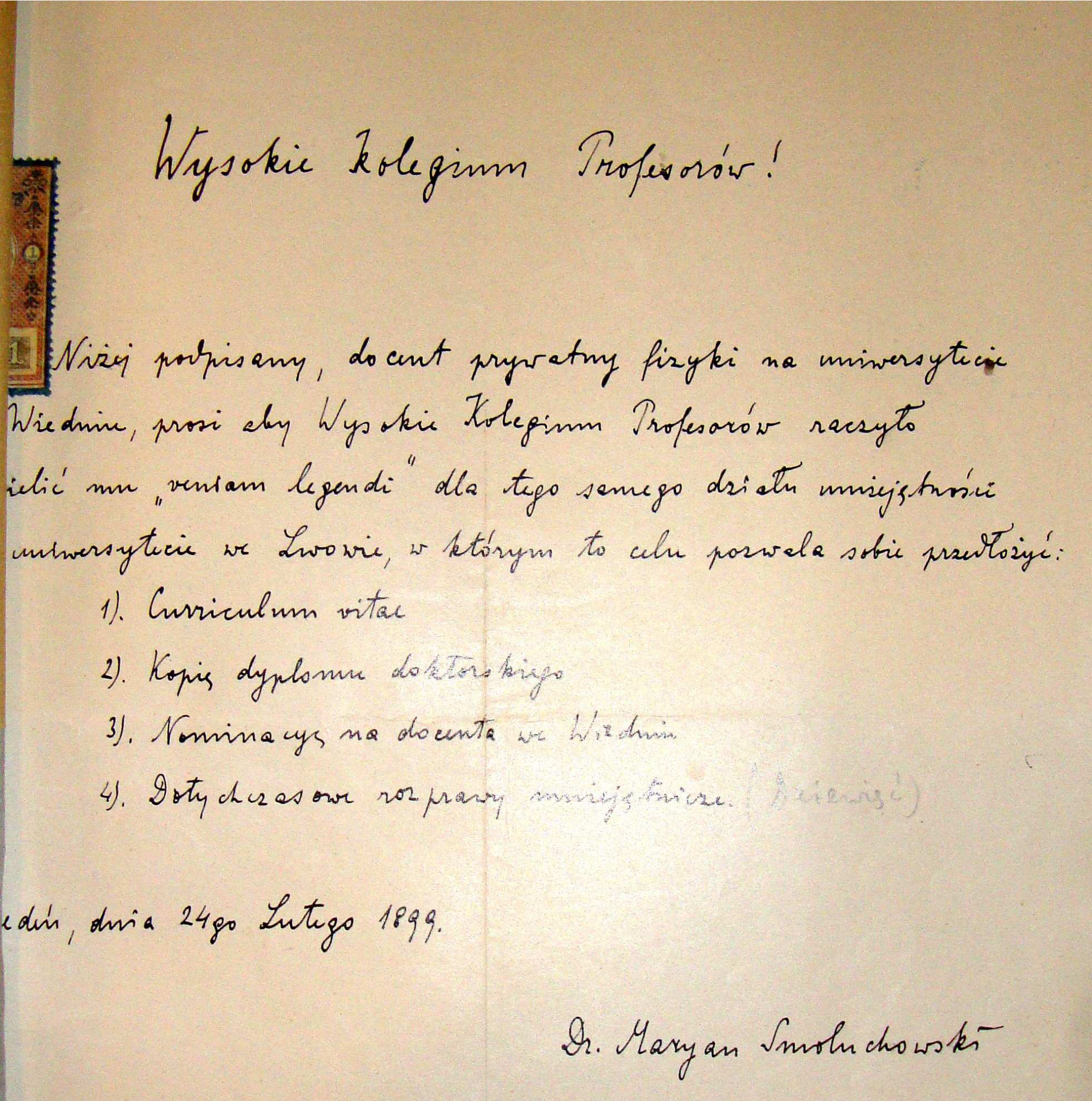} &
\parbox[b]{0.51\textwidth}{%

\it
\hspace*{1cm}Wysokie Kolegium Profesor\'ow!\\
\\
Ni\.zej podpisany, docent prywatny fizyki na uniwersytecie w Wiedniu, prosi aby Wysokie Kolegium Profesor\'ow raczy\l{}o udzieli\'c mu ,,veniam legendi'' dla tego samego dzia\l{}u umiej\k{e}tno\'sci na uniwersytecie we Lwowie, w kt\'orym to celu pozwala sobie przed\l{}o\.zy\'c:\\
\\
1). Curriculum vitae\\
2). Kopi\k{e} dyplomu doktorskiego\\
3). Nominacy\k{e} na docenta we Wiedniu\\
4). Dotychczasowe rozprawy umiej\k{e}tnicze.\\
\\
Wiede\'n, dnia 24go Lutego 1899.\\
\hspace*{3cm}	Dr. Maryan Smoluchowski\\
}
\end{tabular}
\end{table}

The translation of this request is approximately as follows:

\bigskip
\centerline{
\begin{minipage}{0.7\textwidth}
\it
\hspace*{2cm}High Collegium of Professors!\\
\\
The undersigned, a private docent of physics at the University of Vienna, asks that the High Collegium of Professors would give him ``veniam legendi'' for the same skills at the university department in Lviv, for which purpose he takes the liberty to submit:\\
\\
1). Curriculum vitae\\
2). A copy of the doctoral degree diploma\\
3). Appointments to associate professor in Vienna\\
4). Previous scientific works.\\
\\
Vienna, the 24th day of February 1899.\\
\hspace*{6.2cm}Dr. Maryan Smoluchowski
\end{minipage}
}

\bigskip
The request was soon approved and in May, 1899 Marian Smoluchowski was appointed a private docent (lecturer) of theoretical physics at the University of Lviv. Next year (1900) he became an extraordinary (associate) professor, and in 1903 an ordinary (full) professor of theoretical physics.

In 1901 he became Doctor \textit{honoris causa} of law (LLD) at the Glasgow University (in England, honorary doctorate is given in a field different from awardee's primary one).
The period between August, 1905 and April, 1906 Smoluchowski spent at the Cavendish Laboratory in Cambridge doing research with J.J.~Thomson.

Soon after returning to Lviv, he was elected the Dean of the Philosophy Faculty for the academic year 1906/07. In years 1906--08 Smoluchowski was President of the Copernicus Society of Natural Scientists in Lviv (Towarzystwo przyrodnik\'ow imienia Kopernika we Lwowie), where he was a member of the Board since 1900.

In 1908 he was elected a corresponding member of the Polish Academy of Skills in Krak\'ow (in Polish: Akademia Umiej\k{e}tno\'sci, which is also translated into English as Academy of Sciences and Letters), becoming a full member in 1917. Also in 1908 Smoluchowski was awarded the Haitinger prize of the Vienna Academy of Sciences for the theoretical explanation of the Brownian motion. In 1909 he was appointed a member of The Imperial and Royal Commission for Weights and Measures (k.~und~k. Normal-Eichungs-Kommission) in Vienna \cite[p.~524]{Kronika12}.

\begin{wrapfigure}{r}{0.35\textwidth}
\vspace*{-0.50cm}
\centerline{\includegraphics[clip,width=0.32\textwidth]{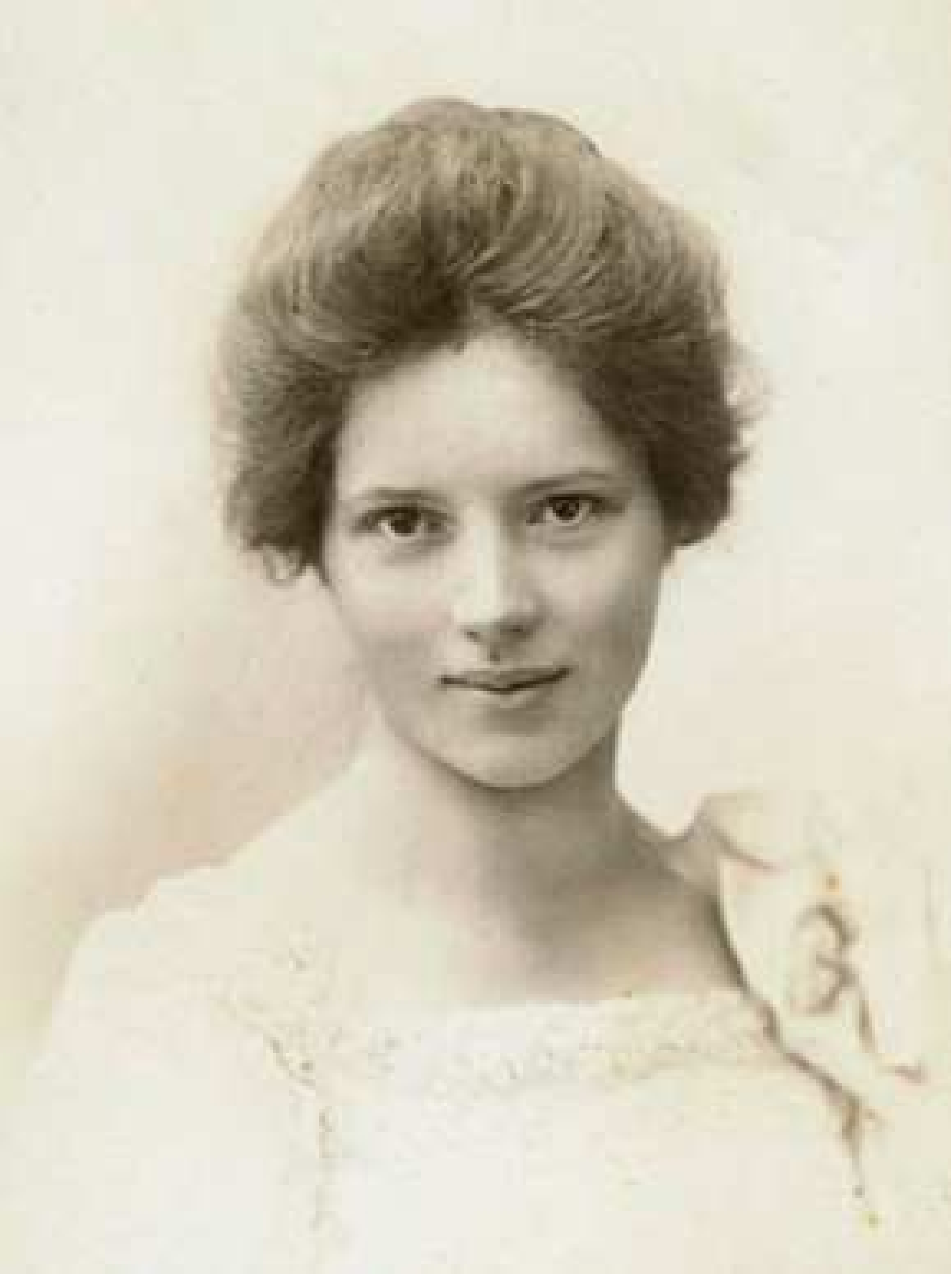}}
\caption{Zofia Baraniecka. Photo by J. Sebald (Krak\'ow, ca. 1901).}
\label{fig:Zofia}
\vspace*{-0.80cm}
\end{wrapfigure}
Finally, in May 1913 Marian Smoluchowski moved to Krak\'ow. His position at the University of Lviv was taken by Konstanty Zakrzewski.

On 01 June 1901 Marian Smoluchowski married Zofia Baraniecka (1881--1959), a daughter of Marian Baraniecki, a mathematics professor of the Jagiellonian University. They had two children: daughter Aldona (1902--1984) and son Roman (1910--1996), who also became a physicist, working as a professor at the universities of Princeton and Austin (USA).

\begin{figure}[!b]
\begin{minipage}{0.49\textwidth}
\centerline{\includegraphics[height=2.5cm]{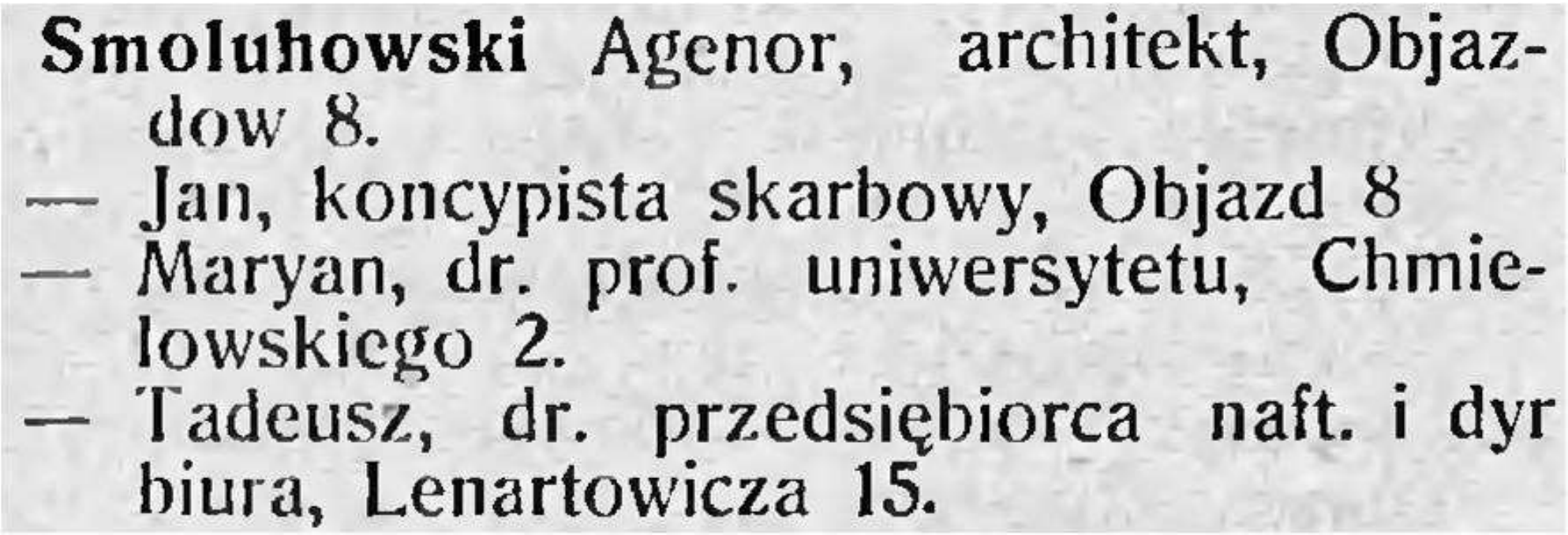}}
\medskip
\centerline{(Year 1910)}
\vspace*{1.cm}
\centerline{\includegraphics[height=2.5cm]{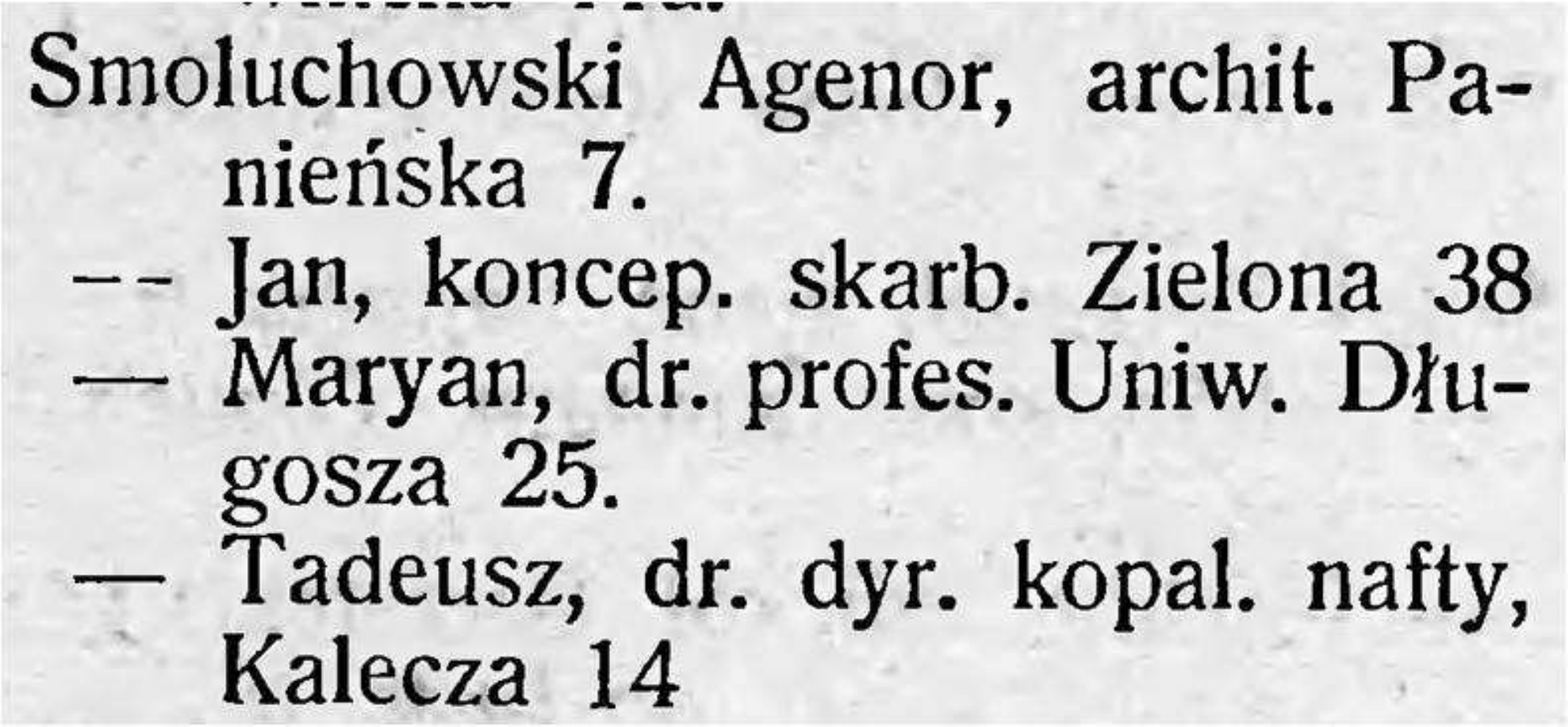}}
\medskip
\centerline{(Year 1913)}
\end{minipage}\quad%
\begin{minipage}{0.49\textwidth}
\centerline{\includegraphics[width=0.96\textwidth]{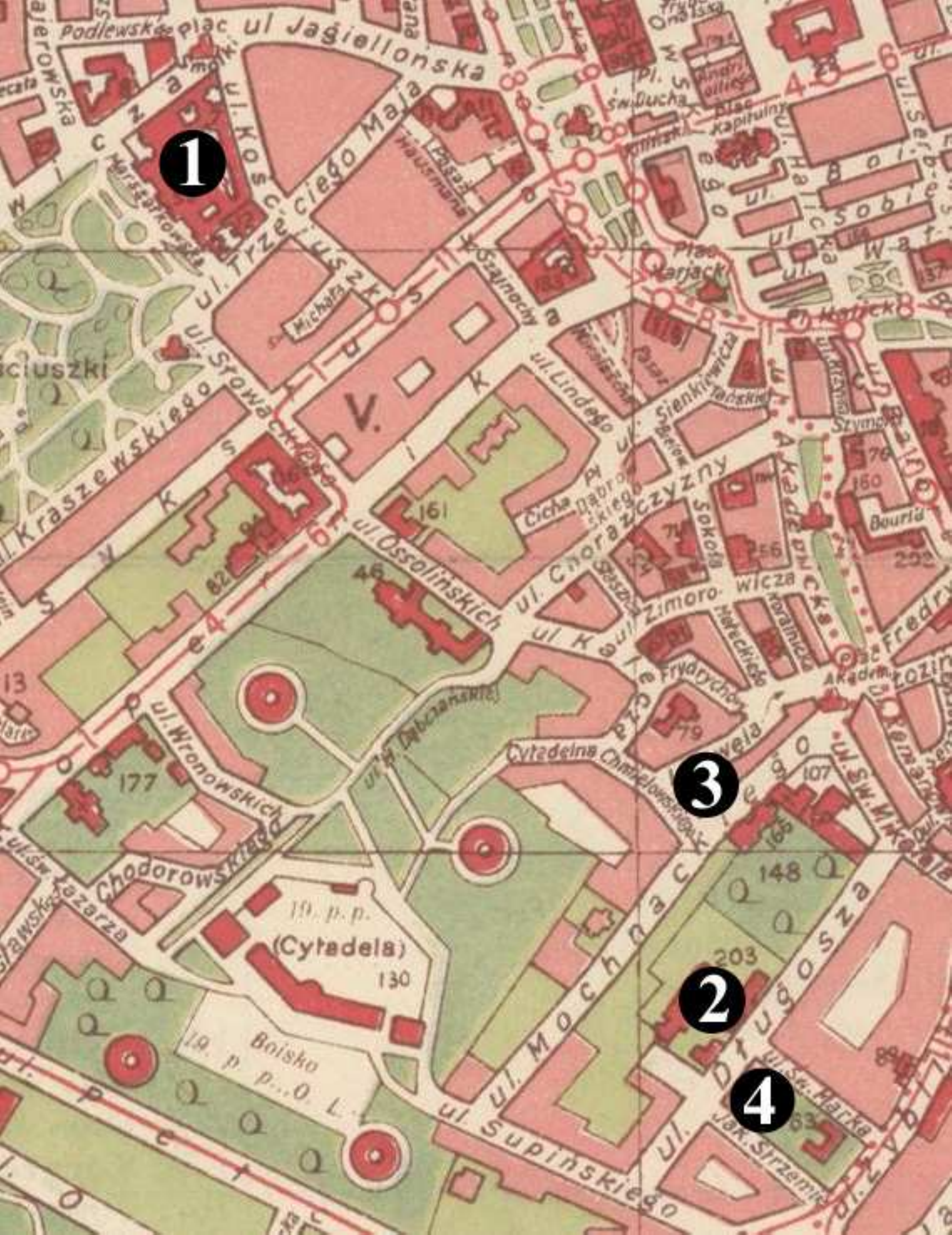}}
\end{minipage}
\medskip
\caption{\textit{Left:} addresses as given in Lviv address books \cite{AddressBook:1910} and \cite{AddressBook:1913}.
\textit{Right:} Map of the central part of Lviv \cite{Map:1929}.
{\sl 1} --- present-time main building of the National University of Lviv (1, Universytetska Street, formerly Marsza\l{}kowska Street),
{\sl 2} --- Institute of Physics (now Faculty of Physics);
{\sl 3} --- Smoluchowskis' address in 1910;
{\sl 4} --- Smoluchowskis' address in 1913.
}
\label{fig:map}
\end{figure}

Two addresses where the Smoluchowskis lived in Lviv can be found in address books of that period~\cite{AddressBook:1910,AddressBook:1913}, see figure~\ref{fig:map}. The one from the year 1910 lists ``Smoluchowski Maryan, dr. prof. uniwersytetu'' under 2,~Chmielowskiego Street (now Hlibova Street); the book from 1913 gives another address: 25,~D\l{}ugosza Street (now Kyryla i Mefodiya Street), nearly in front of the building of the Institute of Physics (now the main building of the Faculty of Physics at 8, Kyryla i Mefodiya Street), see map in figure~\ref{fig:map}.

\begin{figure}[h]
\centerline{\includegraphics[width=0.75\textwidth]{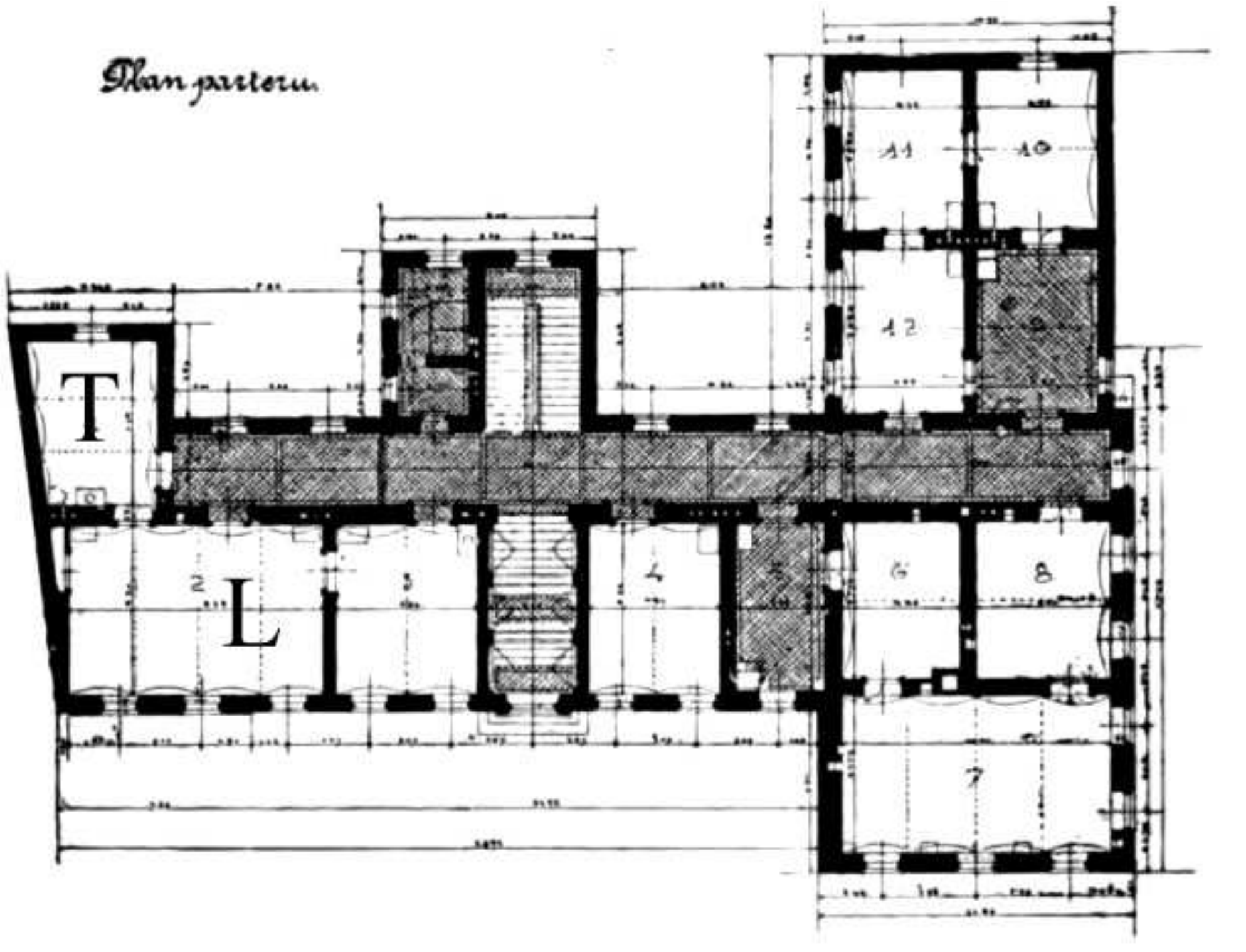}}
\caption{Department location. ``T'' marks the room for professors of mathematics and theoretical physics, ``L'' marks the small lecture hall for students of mathematics, theoretical physics, and geography.}
\label{fig:dept}
\end{figure}

\clearpage
The Department for Theoretical Physics at that time was situated in the newly built (1897) Institute of Physics.
The cabinet for professors of mathematics and theoretical physics was located on the ground floor, next door to the small lecture hall for students of mathematics, theoretical physics, and geography \cite[p.~241--42]{Kronika99}, see figure~\ref{fig:dept}.

\medskip
Lviv period was the most productive in the scientific work of Marian Smoluchowski. This topic is well described and analyzed in the literature, see for instance~\cite{Teske55,Teske77,Teske68,Sredniawa85,Chandr00,Loria53,Fulinski98}, and does not require additional commentaries. Only a brief account of the breadth of his academic interests is thus given.
Marian Smoluchowski worked in various domains: kinetic theory, in particular Brownian motion; phenomena of kataphoresis, electrophoresis, and coagulation; hydrodynamics of viscous liquids; properties of atmospheres of Earth and planets; formation of mountains; etc. More insight about the subjects of his work can be obtained upon analyzing the list of publications from the Lviv period given in section~\ref{sec:publ}.

\section{Teaching}
Marian Smoluchowski brought new teaching methods to the University of Lviv by introducing practical trainings in theoretical physics (two hours per week)~\cite[p.~130]{Teske55}, according to the teaching experience at the University of Vienna and other leading European universities.

During his work as a professor Marian Smoluchowski's main teaching subjects where: \textit{Electricity and Magnetism}, \textit{Mechanics}, and \textit{Thermodynamics}. A complete list of courses he taught is given in table~\ref{table:Courses}.

\begin{figure}[h]
\centerline{%
\includegraphics[width=0.45\textwidth]{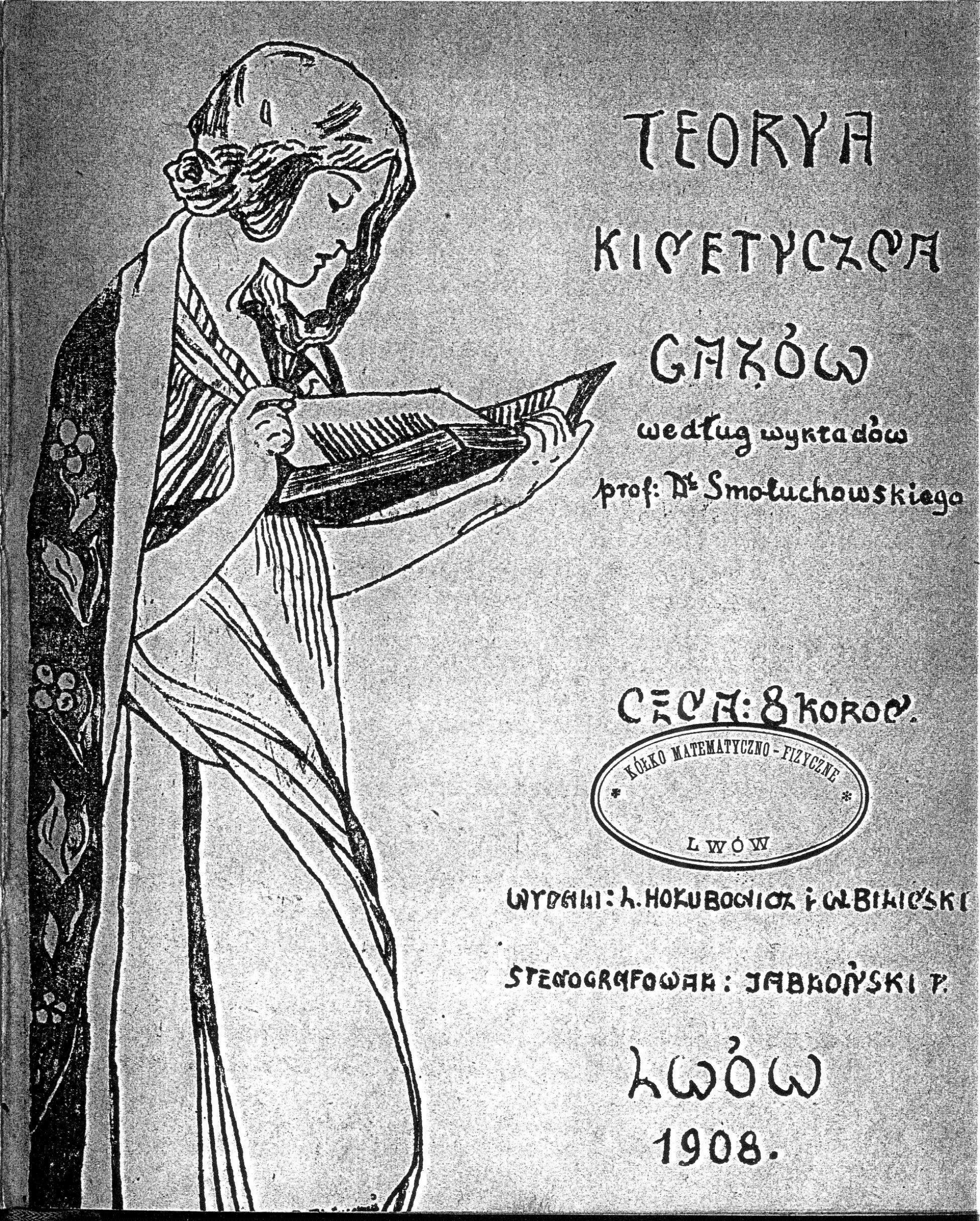}
\quad
\includegraphics[width=0.45\textwidth]{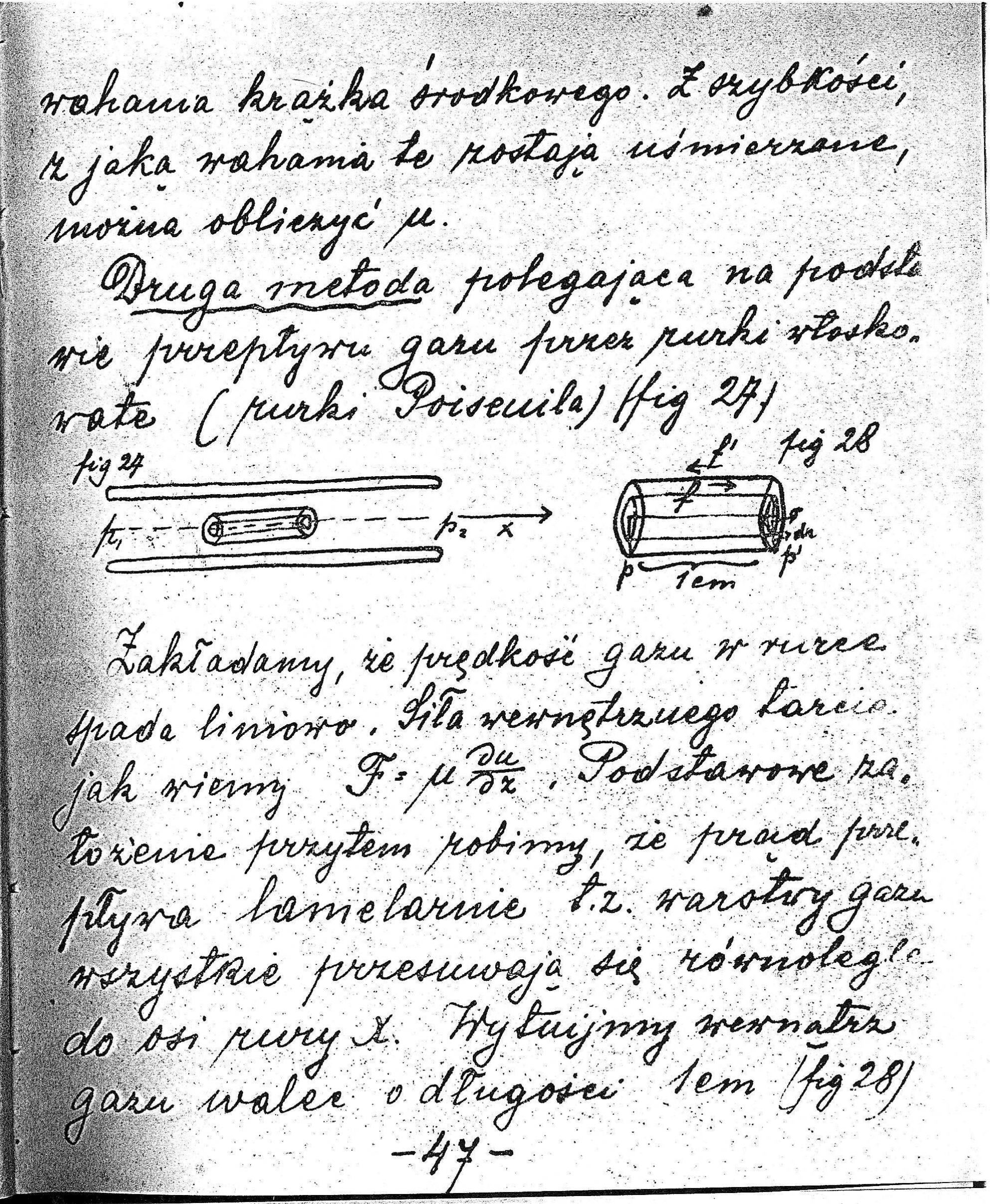}
}
\vspace{1.5cm}
\centerline{%
\includegraphics[width=0.45\textwidth]{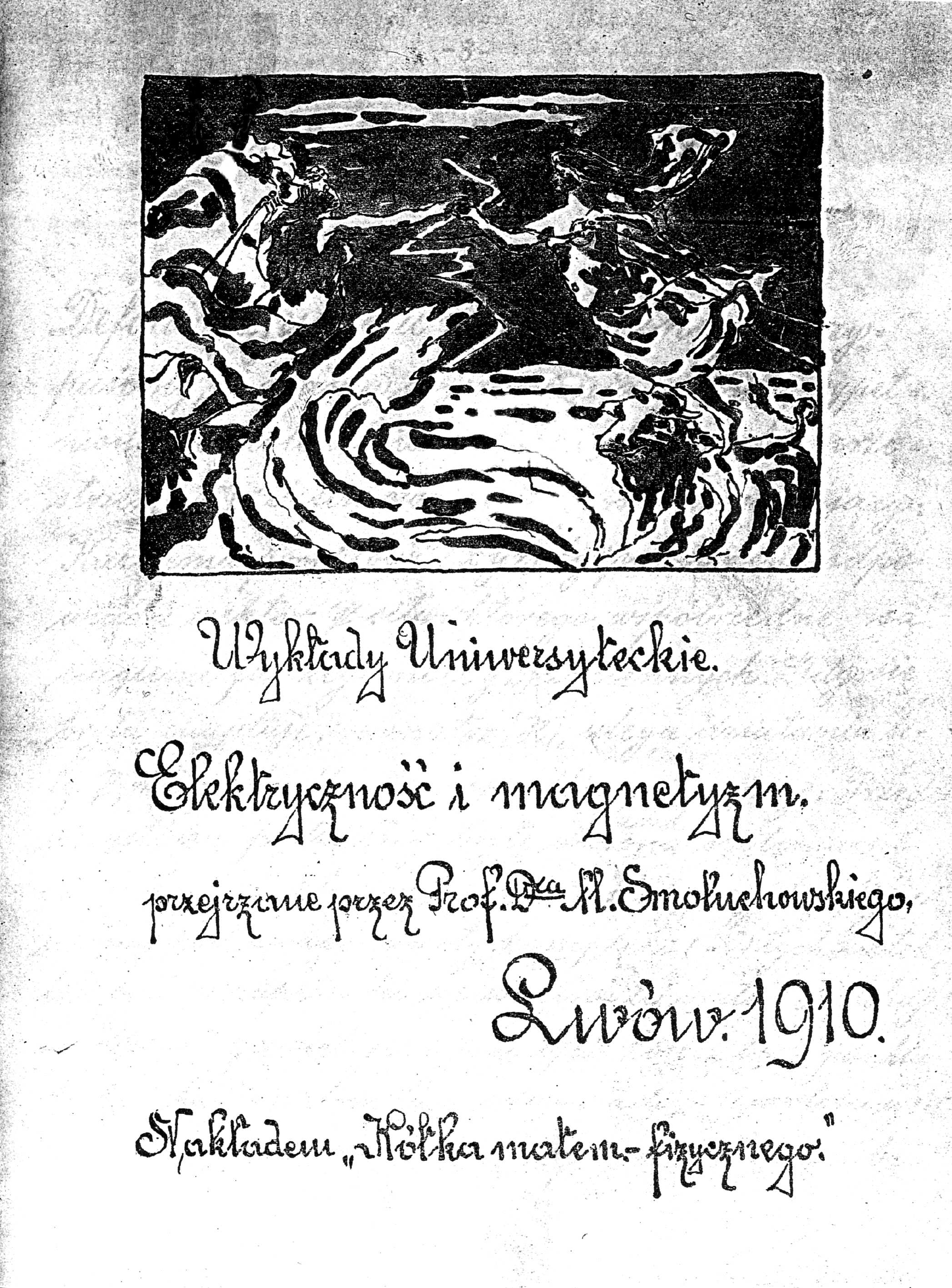}
\quad
\includegraphics[width=0.45\textwidth]{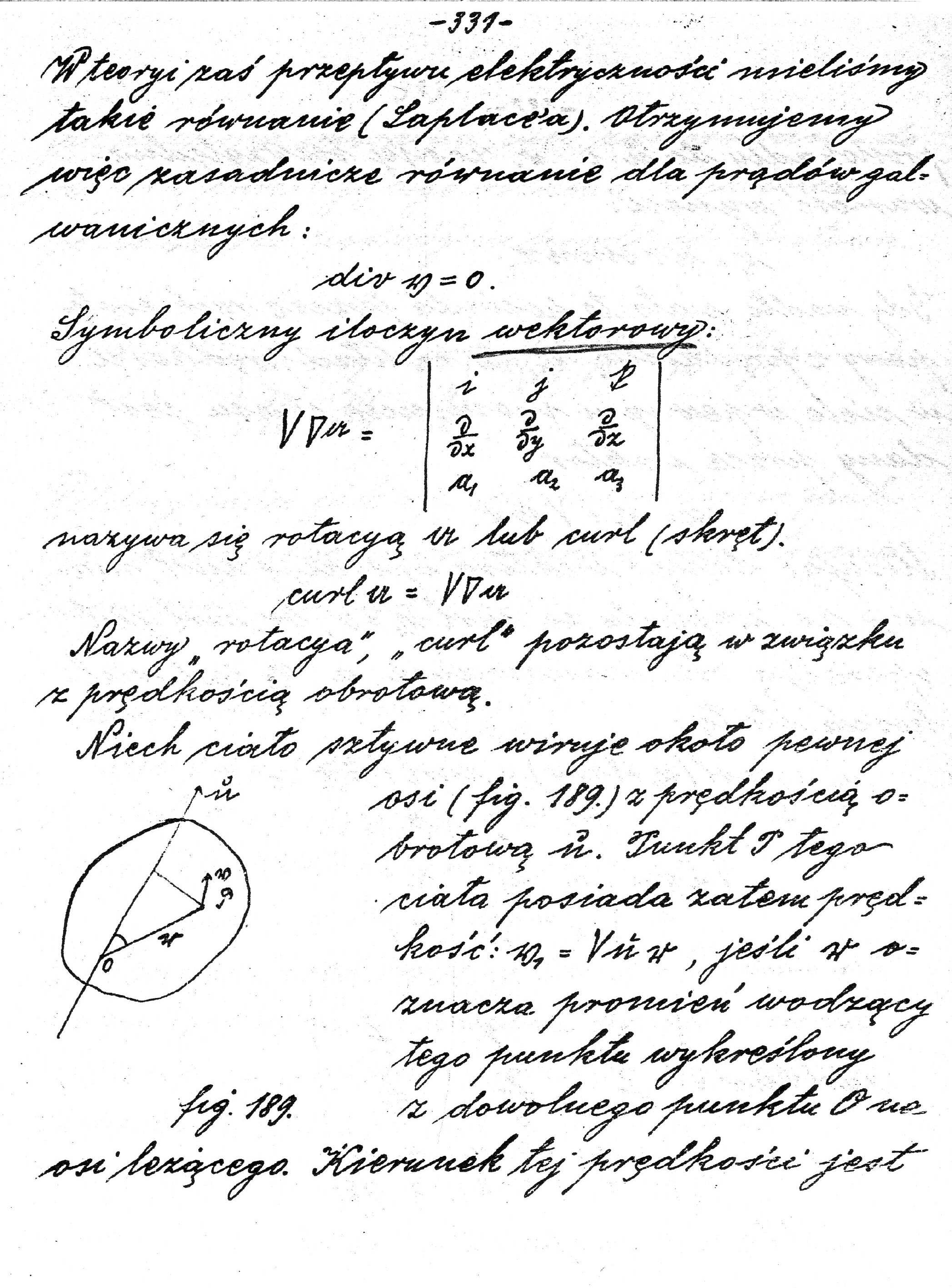}
}
\vspace{5mm}
\caption{Textbooks based on Marian Smoluchowski's lectures and approved by him, published by ``K\'o\l{}ko matematyczno-fizyczne'' (Mathematical-Physical Circle):
\textit{Teorya kinetyczna gaz\'ow [Kinetic theory of gases]} (upper panel) and
\textit{Elektryczno\'s\'c i magnetyzm [Electricity and magnetism]} (lower panel).
}
\label{fig:books}
\end{figure}

Curiously enough, there was a practice in the University of Lviv, at least among students of physics and mathematics, to publish texts of lectures (upon approval of a lecturer). These books were hand-written and illustrated, then self-published by the Mathematical-Physical Circle (K\'o\l{}ko matematyczno-fizyczne in Polish), presumably in several dozens of copies. The technique of anastatic printing (lithography) was used for copying, see figure~\ref{fig:books}.

In 1917, Marian Smoluchowski wrote a major part of \textit{Poradnik dla samouk\'ow [Guidebook for Self-Instruction]}, which was also co-authored by Maurycy Pius Rudzki and Romuald Merecki.

\begin{table}[h]
\caption{Courses by Marian Smoluchowski. Teaching language was Polish. In the last column, the position is mentioned: ``priv. Doc.'' is ``Privatedozent'', ``e.-o. Prof.'' means ``extraordinary Professor'', and ``o.~Prof.'' means ``ordinary Professor''.
($^*$) In summer semester of academic years 1910/1911 and 1911/1912 additional courses on theoretical physics were taught by Dr. Alfred Denizot, Privadozent of Lviv Polytechnics: \textit{Selected topics of analytical mechanics}
and \textit{Refraction of light}, respectively.
}
\vspace{2ex}
\centering
\begin{tabular}{|c|l|c|}
\hline
\textbf{Semester}	&	\textbf{Courses}	&	\textbf{Position}	\\
\hline\hline
1898/1899 (summer)	&	Solid state mechanics	&	priv. Doc.	\\
\hline
1899/1900 (winter)	&	Theory of potential	&	priv. Doc.	\\
\hline
1900/1901 (winter)	&	Electricity and magnetism	&	e.-o. Prof.	\\
	&	Differential calculus	&		\\
\hline
1901/1902 (winter)	&	Thermodynamics	&	e.-o. Prof.	\\
	&	Practical trainings in theoretical physics	&		\\
\hline
1902/1903 (winter)	&	Mechanics	&	e.-o. Prof.	\\
	&	Practical trainings in theoretical physics	&		\\
\hline
1903/1904 (winter)	&	Electricity and magnetism	&	e.-o. Prof.	\\
	&	Practical trainings in theoretical physics	&		\\
\hline
1904/1905 (winter)	&	Thermodynamics	&	o. Prof.	\\
	&	Practical trainings in theoretical physics	&		\\
	&	Spectral analysis and theory of radiation	&		\\
\hline
1905/1906 (winter)	&	Mechanics	&	o. Prof.	\\
	&	Practical trainings in theoretical physics	&		\\
\hline
1906/1907 (winter)	&	Electricity and magnetism	&	o. Prof.	\\
	&	Practical trainings in theoretical physics	&		\\
\hline
1906/1907 (summer)	&	Optics and theory of electrons	&	o. Prof.	\\
	&	Practical trainings in theoretical physics	&		\\
\hline
1907/1908 (winter)	&	Thermodynamics	&	o. Prof.	\\
	&	Practical trainings in theoretical physics	&		\\
\hline
1907/1908 (summer)	&	Kinetic theory of gases	&	o. Prof.	\\
	&	Practical trainings in theoretical physics	&		\\
\hline
1908/1909 (winter)	&	Mechanics	&	o. Prof.	\\
	&	Practical trainings in theoretical physics	&		\\
\hline
1908/1909 (summer)	&	Hydrodynamics and acoustics	&	o. Prof.	\\
	&	Practical trainings in theoretical physics	&		\\
\hline
1909/1910 (winter)	&	Electricity and magnetism	&	o. Prof.	\\
	&	Introduction to theoretical physics	&		\\
	&	Practical trainings in theoretical physics	&		\\
\hline
1909/1910 (summer)	&	New theories of electricity and magnetism	&	o. Prof.	\\
	&	Practical trainings in theoretical physics	&		\\
\hline
1910/1911 (winter)	&	Thermodynamics	&	o. Prof.	\\
	&	Practical trainings in theoretical physics	&		\\
	&	Selected topics of analytical mechanics	&		\\
\hline
1910/1911 (summer)$^*$	&	Kinetic theory of matter	&	o. Prof.	\\
	&	Electric conductivity of gases	&		\\
	&	Practical trainings in theoretical physics	&		\\
\hline
1911/1912 (summer)$^*$	&	Hydrodynamics and acoustics	&	o. Prof.	\\
	&	Selected topics of kinetic theory	&		\\
	&	Tutorial of theoretical physics	&		\\
\hline
1912/1913 (winter)	&	Electricity and magnetism	&	o. Prof.	\\
	&	Tutorial of mathematical physics	&		\\
\hline
1912/1913 (summer)	&	Electromagnetic waves and theory of electrons	&	o. Prof.	\\
	&	Tutorial of mathematical physics	&		\\
\hline
\end{tabular}
\label{table:Courses}
\end{table}


\clearpage

\section{Smoluchowski's publications from the Lviv period}
\label{sec:publ}

The following list is, to my knowledge, the most complete information on Marian Smoluchowski's publications during his stay in Lviv.
The only doubt concerns item~\ref{Taternik:1913}, which can already belong to the Krak\'ow period (no affiliation is mentioned in this paper), while the trip described therein occurred still during the Lviv period.
The detailed information (being updated as necessary) is available~\cite{new}.

Positions marked with an asterisk ($^*$) were not checked \textit{de visu}.

\begin{enumerate}
\def\No{N$^\circ\,$}

\item[\textbf{1900}]

\item \textrm{O atmosferze Ziemi i planet [On the atmosphere of Earth and planets]}. \textit{Ksi\k{e}ga pami\k{a}tkowa wydana przez Uniwersytet lwowski ku uczczeniu 500-letniego jubileuszu Uniwersytetu Jagiello\'nskiego} (Lw\'ow, 1900), 28~s.

\item \textrm{O wynikach nowszych bada\'n nad promieniowaniem [On the results of recent studies of radiation]}. \textit{Kosmos}. 1900. \textbf{25}, S.~74--87.

\item \textrm{[Review:] M. P. Rudzki: O kszta\l{}cie fali spr\k{e}\.zystej w pok\l{}adach ziemskich. (Rozpr. Akad. Um. w Krakowie tom XXXIX.)}. \textit{Kosmos}. 1900. \textbf{25}, S.~671--672.

\item[\textbf{1901}]

\item \textrm{\"Uber die Atmosph\"are der Erde und der Planeten [On the atmosphere of Earth and planets]}. \textit{Phys. Zs.} 1901. \textbf{2}, S.~307--313.

\item \textrm{O nowszych post\k{e}pach na polu teoryj kinetycznych materyi [On recent progress in the field of the kinetic theory of matter]}. \textit{Prace matematyczno-fizyczne}. 1901. \textbf{22}, S.~112--135.

\item \textrm{Kongres mi\k{e}dzynarodowy fizyk\'ow, odbyty w Pary\.zu od d. 6-12 sierpnia 1900 r. [International congress of physicists held in Paris on 6-12 August 1900]} \textit{Wiadomo\'sci matematyczne}. 1901. \textbf{5}, S.~80--89.

\item \textrm{[Review:] J. Zawidzki. Ueber die Dampfdrucke bin\"arer Fl\"ussigkeitgemische. (Inaugural Dissertation, Leipzig, 1900)}. \textit{Kosmos}. 1901. \textbf{26}, S.~48.

\item \textrm{[Review:] L. Silberstein. Symbolische Integrale der elektromagnetischen Gleichungen, aus dem Anfangszustand des Feldes abgeleitet, nebst Andeutungen zu einer allgemeinen Theorie physikalischer Operatoren. (Annalin d. Physik 6 p. 373 (2902)}. \textit{Kosmos}. 1901. \textbf{26}, S.~565.

\item[\textbf{1902}]

\item \textrm{[Review:] Alfred Denizot. Ueber ein Pendelproblem von Euler. (Verhandl. Deutsch. Phys. Gesellsch. III. p. 213.)}. \textit{Kosmos}. 1902. \textbf{27}, S.~32.

\item \textrm{[Review:] W\l{}adys\l{}aw Natanson: 1. Pogl\k{a}d na rodzaje zjawisk [\ldots] 7. O podw\'ojnem za\l{}amaniu \ldots}. \textit{Kosmos}. 1902. \textbf{27}, S.~240--243.

\item \textrm{[Review:] C. Zakrzewski: O oscylacyi kr\k{a}\.zka w p\l{}ynie lepkim. (Sur les oscillations d'un disque plong\'e dans un liquide visqueux). (Odb. Bullet. Acad. Crac. 1902, p. 239)}. \textit{Kosmos}. 1902. \textbf{27}, S.~243--244.

\item \textrm{[Review:] Wac\l{}aw Wolski: O taranie wiertniczym. (Lw\'ow 1902, Drukarnia ,,S\l{}owa Polskiego``}. \textit{Kosmos}. 1902. \textbf{27}, S.~244.

\item[\textbf{1903}]

\item \textrm{O zjawiskach aerodynamicznych i po\l{}\k{a}czonych z niemi objawach cieplnych [On the aerodynamic phenomena and the associated thermal effects]}. \textit{Rozprawy Wydzia\l{}u mat.-przyrod. Akad. Um.} 1903. \textbf{3A\,(43A)}, S.~71--109.

\item \textrm{Sur les ph\'enom\`enes a\'erodynamiques et les effets thermiques qui les accompagnent [\textsl{ut supra}]}. \textit{Bull. Int. Acad. Sci. Cracovie. Classe des sciences math. et naturelles}. 1903. P.~143--182.

\item \textrm{Przyczynek do teoryi endosmozy elektrycznej i kilku zjawisk pokrewnych [Contribution to the theory of electrical endosmosis and related phenomena]}. \textit{Rozprawy Wydzia\l{}u mat.-przyrod. Akad. Um.} 1903. \textbf{3A\,(43A)}, S.~110--127.

\item \textrm{Contribution \`a la th\'eorie de l'endosmose \'electrique et de quelques ph\'enom\`enes corr\'elatifs \, \linebreak{} [\textsl{ut supra}]}. \textit{Bull. Int. Acad. Sci. Cracovie. Classe des sciences math. et naturelles}. 1903. P.~182--199.

\item[\textbf{1904}]

\item \textrm{On the principles of aerodynamics and their application, by the method of dynamical similarity, to some special problems}. \textit{Phil. Mag. Series 6}. 1904. \textbf{7}, P.~667--681.

\item \textrm{O metodzie podobie\'nstwa dynamicznego i jej zastosowaniach w mechanice cieczy i gaz\'ow [On the method of dynamic similarity and its application to the mechanics of liquids and gases]}. \textit{Prace matematyczno-fizyczne}. 1904. \textbf{15}, S.~115--134.

\item \textrm{\"Uber Unregelm\"a\ss{}igkeiten in der Verteilung von Gasmolek\"ulen und deren Einflu\ss{} auf Entropie und Zustandsgleichung [On irregularities of the distribution of molecules in gases and their influence on entropy and the equation of state]}. \textit{Festschrift Ludwig Boltzmann gewidmet zum sechzigsten Geburtstage 20. Februar 1904} (Leipzig: Verlag von Johann Ambrosius Barth, 1904), S.~626--641.

\item \textrm{O powstawaniu \.zy\l{} podczas wyp\l{}ywu cieczy [On the formation of veins in the flow of liquids]}. \textit{Rozprawy Wydzia\l{}u mat.-przyrod. Akad. Um.} 1904. \textbf{4A\,(44A)}, S.~144--157.

\item \textrm{Sur la formation des veines d'efflux dans les liquides [\textsl{ut supra}]}. \textit{Bull. Int. Acad. Sci. Cracovie. Classe des sciences math. et naturelles}. 1904. P.~371--384.

\item \textrm{Sprawozdania z prac polskich na polu fizyki za lata 1901 i 1902 [Reports on Polish works in the field of physics in the years 1902 and 1902]}. \textit{Kosmos}. 1904. \textbf{29}, S.~528--545.

\item \textrm{[Review:] Stronhal \v{C}. Mechanika. (670 str. 342 ilustr.) w Praze 1901; Akustika (462 str. 144 ilustr.) w Praze 1903)}. \textit{Kosmos}. 1904. \textbf{29}, S.~551--552.

\item[\textbf{1905}]

\item \textrm{Zur Theorie der elektrischen Kataphorese und der Oberfl\"achenleitung [On the theory of electric cataphoresis and surface conduction]}. \textit{Phys. Zs.} 1905. \textbf{6}, S.~529--531.

\item[\textbf{1906}]

\item \textrm{Zur kinetischen Theorie der Brownschen Molekularbewegung und der Suspensionen [To the kinetic theory of Brownian motion and the suspensions]}. \textit{Ann. d. Phys.} 1906. \textbf{21\,(326)}, S.~756--780.

\item \textrm{O drodze \'sredniej cz\k{a}steczek gazu i o zwi\k{a}zku jej z teory\k{a} dyfuzyi [On the mean free path of gaseous molecules and its relation to the theory of diffusion]}. \textit{Rozprawy Wydzia\l{}u mat.-przyrod. Akad. Um.} 1906. \textbf{6A\,(46A)}, S.~129--139.

\item \textrm{Sur le chemin moyen parcouru par les mol\'ecules d'un gaz et sur son rapport ayec la th\'eorie de la diffusion [\textsl{ut supra}]}. \textit{Bull. Int. Acad. Sci. Cracovie. Classe des sciences math. et naturelles}. 1906. P.~202--213.

\item \textrm{Zarys teoryi kinetycznej ruch\'ow Browna i roztwor\'ow m\k{e}tnych [Outline of the kinetic theory of Brownian motion and turbid media]}. \textit{Rozprawy Wydzia\l{}u mat.-przyrod. Akad. Um.} 1906. \textbf{6A\,(46A)}, S.~257--281.

\item \textrm{Essai d'une th\'eorie cin\'etique du muovement Brownien et des milieux troubles [\textsl{ut supra}]}. \textit{Bull. Int. Acad. Sci. Cracovie. Classe des sciences math. et naturelles}. 1906. P.~577--602.

\item[\textbf{1907}]

\item \textrm{Przyczynek do teoryi ruch\'ow cieczy lepkich, zw\l{}aszcza zagadnie\'n dwuwymiarowych [Contribution to the theory of the motion of viscous liquids, especially in two-dimensional problems]}. \linebreak \textit{Roz\-prawy Wydzia\l{}u mat.-przyrod. Akad. Um.} 1907. \textbf{7A\,(47A)}, S.~1--16.

\item $^*$ \textrm{Contribution \`a la th\'eorie du mouvement des liquides visqueux; en particulier des probl\`emes en deux dimensions [\textsl{ut supra}]}. \textit{Bull. Int. Acad. Sci. Cracovie. Classe des sciences math. et naturelles}. 1907. P.~1--16.

\item \textrm{Teorya kinetyczna opalescencyi gaz\'ow w stanie krytycznym oraz innych zjawisk pokrewnych [Kinetic theory of opalescence of gases in the critical state and other related phenomena]}. \textit{Rozprawy Wydzia\l{}u mat.-przyrod. Akad. Um.} 1907. \textbf{7A\,(47A)}, S.~179--198.

\item $^*$ \textrm{Th\'eorie cin\'etique de l'opalescence des gaz \`a l'\'etat critiqueet de certains ph\'enomen\`es correlatifs [\textsl{ut supra}]}. \textit{Bull. Int. Acad. Sci. Cracovie. Classe des sciences math. et naturelles}. 1907. P.~1057--1075.

\item \textrm{Zarys najnowszych post\k{e}p\'ow fizyki [An outline of the latest advances in physics]}. \textit{Muzeum: czasopismo po\'swi\k{e}cone sprawom wychowania i szkolnictwa (Lw\'ow)}. 1907. \textbf{23, T.~I}, S.~43-60; 144--165.

\item \textrm{[Introductory speech at the XXXVII Extraordinary General Meeting of the Polish Copernicus Sosiety of Natural Scientists]}. \textit{Kosmos}. 1907. \textbf{32}, S.~255--257.

\item[\textbf{1908}]

\item \textit{Teorya kinetyczna gaz\'ow} / wed\l{}ug wyk\l{}adow prof. Dr Smoluchowskiego [\textit{Kinetic theory of gases} / based on lectures by Prof. Dr. Smoluchowski] (Lw\'ow: wydali: L.~Ho\l{}ubowicz i W.~Bili\'nski; Steno\-grafowa\l{}: Jablo\'nski P., 1908), 213~l.

\item \textrm{Molekular-kinetische Theorie der Opaleszenz von Gasen im kritischen Zustande, sowie einiger verwandter Erscheinungen [Molecular-kinetic theory of opalescence of gases in the critical state and other related phenomena]}. \textit{Ann. d. Phys.} 1908. \textbf{25\,(330)}, S.~205--226.

\item $^*$ \textrm{Uwagi o kilku zjawiskach drobinowych, zwi\k{a}zanych z przypadkowemi odchyleniami od stanu najprawdopodobniejszego [Remarks on some molecular phenomena connected with accidental deviations from the most probable state]}. \textit{Sprawozdanie X-go Zjazdu Lekarzy i Przyrodnik\'ow Polskich} (Lw\'ow, 1908), 19~s.

\item \textrm{Lord Kelvin [: Obituary]}. \textit{Ateneum Polskie}. 1908. \textbf{Nr.~1, T.~I}, S.~212--228.

\item \textrm{[Review:] Dr. W\l{}adys\l{}aw Natanson: ,,Odczyty i Szkice`` Warszawa 1908, Wende i Sp.} \textit{Ateneum Polskie}. 1908. \textbf{Nr.~1, T.~II}, S.~134--136.

\item \textrm{Stanis\l{}aw K\k{e}pi\'nski [: Obituary]}. \textit{Ateneum Polskie}. 1908. \textbf{Nr.~1, T.~II}, S.~274--276.

\item \textrm{[Review:] Dwie ksi\k{a}\.zki z dziedziny ,,filozofji przyrody``. H. Poincar\'e. ,,Nauka i Hypoteza`` (La science et l'hipothese). Przek\l{}ad M. H. Horwitza, pod redakcj\k{a} Ludwika Silbersteina. Warszawa 1908, 198 str. H. Poincar\'e. ,,Warto\'s\'c Nauki`` (La valeur de la science). Przek\l{}ad Ludwika Silbersteina. Warszawa 1908, 178 str.} \textit{Ateneum Polskie}. 1908. \textbf{Nr.~1, T.~IV}, S.~291--296.

\item \textrm{[Introductory speech at the XXXVII General Meeting of the Polish Copernicus Sosiety of Natural Scientists]}. \textit{Kosmos}. 1908. \textbf{33}, S.~87--94.

\pagebreak
\item[\textbf{1909}]

\item \textrm{Some remarks on the mechanics of overthrusts}. \textit{Geological Mag. New Ser., Decade V}. 1909. \textbf{6}, P.~204--205.

\item \textrm{O pewnem zagadnieniu z teoryi spr\k{e}\.zysto\'sci i o zwi\k{a}zku jego z wytworzeniem si\k{e} g\'or fa\l{}dowych [On a certain problem in the theory of elasticity and its relation to the formation of overthrusts]}. \textit{Rozprawy Wydzia\l{}u mat.-przyrod. Akad. Um.} 1909. \textbf{9A\,(49A)}, S.~223--226.

\item $^*$ \textrm{\"Uber ein gewisses Stabilit\"atsproblem der Elastizit\"atslehre und dessen Beziehung zur Entstehung von Faltengebirgen [\textsl{ut supra}]}. \textit{Bull. Int. Acad. Sci. Cracovie. Classe des sciences math. et naturelles}. 1909. \textbf{II}, P.~3--20.

\item $^*$ \textrm{Versuche \"uber Faltungserscheinungen schwimmender elastischer Platten [Experiments on overthrusts in floating elastic plates]}. \textit{Bull. Int. Acad. Sci. Cracovie. Classe des sciences math. et naturelles}. 1909. \textbf{II}, P.~727--734.

\item \textrm{Kilka uwag o fizycznych podstawach teoryi g\'orotw\'orczych [Several notes on the physical foundations of tectonic theories]}. \textit{Kosmos}. 1909. \textbf{34}, S.~547--579.

\item \textrm{[Review:] L. Brunner. --- Ewolucya Materyi. Zarys nauki o promieniowaniu. [Str. 140. Krak\'ow, (1909). Sk\l{}ad u Gebethnera i u Wendego. Cena 3 K]}. \textit{Kosmos}. 1909. \textbf{34}, S.~1237--1238.

\item[\textbf{1910}]

\item \textit{Elektryczno\'s\'c i magnetyzm: wyk\l{}ady uniwersyteckie} / przejrzane przez Prof. D-ra M.~Smoluchows\-kie\-go [\textit{Electricity and magnetism: university lectures} / revised by Prof. Dr. M.~Smoluchowski] (Lw\'ow: Nak\l{}adem ,,K\'o\l{}ka matem.-fizycznego``, 1910), 416~s.

\item \textit{Teorya Maxwell'a i teorya elektron\'ow} / wed\l{}ug wyk\l{}ad\'ow Prof. Dra Smoluchows\-kie\-go [\textit{Maxwell's theory and theory of electrons}  / based on lectures by Prof. Dr. Smoluchowski] (Lw\'ow: Nak\l{}adem K\'o\l{}ka matemat.-fizycznego, 1910), 245~s.

\item \textrm{Zur kinetischen Theorie der Transpiration und Diffusion verd\"unnter Gase [To the kinetic theory of transpiration and diffusion of dilute gases]}. \textit{Ann. d. Phys.} 1910. \textbf{33\,(338)}, S.~1559--1570.

\item \textrm{Sur la th\'eorie m\'ecanique de l'\'erosion glaciaire [On the mechanical theory of glacial erosion]}. \linebreak \textit{Comptes Rendus hebdomadaires des s\'eances de l'Acad\'emie des Sciences}. 1910. \textbf{150}, P.~1368--1371.

\item \textrm{O przewodnictwie cieplnem cia\l{} sproszkowanych [On the heat conduction of powders]}. \textit{Rozprawy Wydzia\l{}u mat.-przyrod. Akad. Um.} 1910. \textbf{10A\,(50A)}, S.~83--95.

\item $^*$ \textrm{Sur la conductibilit\'e calorifique des corps pulv\'eris\'es [\textsl{ut supra}]}. \textit{Bull. Int. Acad. Sci. Cracovie. Classe des sciences math. et naturelles}. 1910. P.~129--153.

\item \textrm{Przyczynek do kinetycznej teoryi transpiracyi, dyfuzyi i przewodnictwa cieplnego w gazach roz\-rzedzonych [\textsl{ut infra}]}. \textit{Rozprawy Wydzia\l{}u mat.-przyrod. Akad. Um.} 1910. \textbf{10A\,(50A)}, S.~209--214.

\item $^*$ \textrm{Contributions to the theory of transpiration, diffusion and thermal conduction in rarified gases}. \textit{Bull. Int. Acad. Sci. Cracovie. Classe des sciences math. et naturelles}. 1910. P.~295--312.

\item \textrm{Van der Waalsa teorya stanu ciek\l{}ego a zjawiska lepko\'sci [Van der Waals theory of the liquid state and the phenomena of viscosity]}. \textit{Kosmos}. 1910. \textbf{35}, S.~543--549.

\pagebreak
\item[\textbf{1911}]

\item $^*$ \textit{Termodynamika } / Pod\l{}ug wyk\l{}ad\'ow Prof. M. Smoluchowskiego w Uniwersytecie lwowskim w p\'o\l{}roczu zimowym 1910/11 [\textit{Thermodynamics} / based on the lectures by Prof. M.~Smoluchowski in the University of Lviv in winter semester of 1910/11] (Lw\'ow: Nak\l{}adem K\'o\l{}ka matemat-fizycznego, 1911), 293~s.

\item \textit{Mechanika: wyk\l{}ady uniwersyteckie} / wed\l{}ug wyk\l{}ad\'ow Prof. M. Smoluchowskiego w Uniwersytecie lwowskim roku akd. 1911/12 [\textit{Mechanics: university lectures} / based on the lectures by Prof. M.~Smoluchowski in the University of Lviv in the academic year 1911/12] (Lw\'ow: Nak\l{}adem K\'o\l{}ka matemat-fizycznego, [1911/12]), 358,~III~s.~+~Tytul~nenumerowany~(+2~nlb.).

\item \textrm{Some remarks on conduction of heat through rarefied gases}. \textit{Phil. Mag. Series 6}. 1911. \textbf{21}, P.~11--14.

\item \textrm{Bemerkung zur Theorie des absoluten Manometers von Knudsen [Comments on the theory of Knudsen's absolute manometer]}. \textit{Ann. d. Phys.} 1911. \textbf{34 (339)}, S.~182--184.

\item \textrm{Zur Theorie der W\"armeleitung in verd\"unnten Gasen und der dabei auftretenden Druckkr\"afte [On the theory of heat conduction in rarefied gases and of the resulting pressures]}. \textit{Ann. d. Phys.} 1911. \textbf{35\,(340)}, S.~983--1004.

\item $^*$ \textrm{Zur Theorie der W\"armeleitung in verd\"unnten Gasen und der dabei auftretenden Druekkr\"afte [\textsl{ut supra}]}. \textit{Bull. Int. Acad. Sci. Cracovie. Classe des sciences math. et naturelles}. 1911. P.~432--453.

\item \textrm{\"Uber W\"armeleitung pulverf\"ormiger K\"orper und ein hierauf gegr\"undetes neues W\"arme-Iso\-lier\-ung\-sverfahren [On heat conduction of powders and its relation to a new method of thermal insulation]}. \textit{Bericht \"uber den II. internationalen K\"altekongress, Wien 1910, 6-12 Oktober} (Wien: Verlag des II. Internat. K\"altekongress; Druck von J. Weiner, 1911), Band~II, S.~166--172.

\item \textrm{O oddzia\l{}ywaniu wzajemnem kul poruszaj\k{a}cych si\k{e} w o\'srodku lepkim [On mutual interaction of \ spheres moving in a viscous medium]}. \textit{Rozprawy Wydzia\l{}u mat.-przyrod. Akad. Um.} 1911. \textbf{11A} \textbf{(51A)}, S.~1--3.

\item $^*$ \textrm{\"Uber die Wechselwirkung von Kugeln, die sich in einer z\"ahen Fl\"ussigkeit bewegen [\textsl{ut supra}]}. \textit{Bull. Int. Acad. Sci. Cracovie. Classe des sciences math. et naturelles}. 1911. P.~28--39.

\item $^*$ \textrm{Beitrag zur Theorie der Opaleszenz von Gasen im kritischen Zustande [Contribution to the theory of gaseous opalescence in the critical state]}. \textit{Bull. Int. Acad. Sci. Cracovie. Classe des sciences math. et naturelles}. 1911. P.~493--502.

\item $^*$ \textrm{\'Etudes sur la conductibilit\'e calorifique des corps pulv\'eris\'es (suite)} [Studies of heat conduction of powders (continuation)]. \textit{Bull. Int. Acad. Sci. Cracovie. Classe des sciences math. et naturelles}. 1911. P.~548--557.

\item \textrm{Ewolucya teoryi atomistycznej [Evolution of the atomic theory]}. \textit{Rocznik Akademji Umiej\k{e}tno\'sci w Krakowie}. 1910--1911, S.~131--154;
\textit{Wiadomo\'sci Matematyczne}. 1911. \textbf{15}, S.~201--216.

\item $^*$ \textrm{Atomistyka wsp\'o\l{}czesna [Contemporary atomistics]}. \textit{Ksi\k{e}ga Pami\k{a}tkowa XI-go Zjazdu Lekarzy i Przyrodnik\'ow Polskich w Krakowie, 18--22 lipca 1911}, S.~129--143.

\pagebreak
\textbf{The following are three patents on ``the insulation material for Dewar vessels''}:

\item \textit{W\"armeisolierendes Gef\"a\ss{} mit luftleer gemachten Hohlw\"anden (Dewarsches Gef\"a\ss{})}. Kaiserliches Pa\-tentamt. Patentschrift \No268490. Patentiert im Deutschen Reiche vom 2. Februar 1911 ab.

\item \textit{Mati\`ere calorifuge pour les r\'ecipients Dewar}. R\'epublique Fran\c{c}aise. Brevet d'invention\linebreak \No425.542. D\'elivr\'e le 7 avril 1911.

\item \textit{W\"armeisolierendes Material f\"ur Dewar'sche Gef\"a\ss{}e}. \"Osterreichische Patentschrift Nr. 47771. Ausgegeben am 10. Mai 1911.

\item[\textbf{1912}]

\item \textrm{O pewnem zagadnieniu kinetycznej teoryi roztwor\'ow [On a certain problem of the kinetic theory of solutions]}. \textit{Ksi\k{e}ga Pami\k{a}tkowa ku uczczeniu dw\'ochsetnej pi\k{e}\'cdziesi\k{a}tej rocznicy za\l{}o\.zenia Uniwersytetu Lwowskiego przez Kr\'ola Jana Kazimierza} (Lw\'ow, 1912), {Tom~II}, 8~s.

\item \textrm{Experimentell nachweisbare, der \"ublichen Thermodynamik widersprechende Molekular\-ph\"ano\-mene [Experimentally observable molecular phenomena which contradict conventional thermodynamics]}. \textit{Phys. Zs.} 1912. \textbf{13}, S.~1069--1080.

\item $^*$ \textrm{Experimentell nachweisbare, der \"ublichen Thermodynamik widersprechende Molekular\-ph\"ano\-mene [\textsl{ut supra}]}. \textit{Verhandlungen der Versammlung deutscher Naturforscher und \"Arzte. M\"unster}. 1912. \textbf{2}, S.~83.

\item \textrm{On opalescence of gases in the critical state}. \textit{Phil. Mag. Ser.~6}. 1912. \textbf{23}, P.~165--173.

\item \textrm{Erg\"anzungen zur Stokesschen Formel [Supplements to the Stokes formula; Abstract from the  Fifth International Congress of Mathematicians in Cambridge]}. \textit{Phys. Zs.} 1912. \textbf{13}, S.~1135.

\item $^*$ \textrm{Einige Beispiele Brown'scher Molekularbewegung unter Einfluss \"ausserer Kr\"afte [Some examples of Brownian motion under the influence of external forces]}. \textit{Bull. Int. Acad. Sci. Cracovie. Classe des sciences math. et naturelles}. 1912. P.~418--434.

\item \textrm{W\l{}awys\l{}aw Gosiewski [: Obituary]}. \textit{Kosmos}. 1912. \textbf{37}, S.~205.

\item[\textbf{1913}]

\item \textrm{Anzahl und Gr\"osse der Molek\"ule und Atome [The number and size of molecules and atoms]}. \textit{Scientia: rivista internazionale di sintesi scientifica}. 1913. \textbf{13}, P.~27--44.

\item \textrm{Liczba i wielko\'s\'c cz\k{a}steczek i atom\'ow [\textsl{ut supra}]}. \textit{Wiadomo\'sci Matematyczne}. 1913. \textbf{17}, S.~315--329.

\item \textrm{On the practical applicability of Stokes' law of resistance, and the modifications of it required in certain cases}. \textit{Proceedings of the Fifth International Congress of Mathematicians (Cambridge, 22-28 August 1912)} (Cambridge: at the University Press, 1913) {Vol.~II.~Communications to Sections II-IV}, P.~192--201.

\item \textrm{G\"ultigkeitsgrenzen des zweiten Hauptsatzes der W\"armetheorie [Limits of validity of the second law of thermodynamics]}. \textit{Jahresbericht der Deutschen Mathematiker-Vereinigung}. 1913. \textbf{22}, S.~61--64.

\item \textrm{Dzisiejszy stan teoryi atomistycznej [The present status of atomistic theory]}. \textit{Kosmos}. 1913. \textbf{38}, S.~355--373.

\item \textrm{August Witkowski [: Obituary]}. \textit{Kosmos}. 1913. \textbf{38}, S.~305--308.

\item \textrm{Mihailecul (1926 m) i Farcaul (1961 m) w zimie [Mihailecul and Farcaul in winter]}. \textit{Taternik, organ Sekcji Turystycznej Towarzystwa Tatrza\'nskiego (Krak\'ow)}. 1913. \textbf{7}, S.~103--107.
\label{Taternik:1913}

\clearpage

\item[\textbf{1914}]

\item \textrm{Elektrische Endosmose und Str\"omungsstr\"ome [Electrical endosmosis and streaming current]}. \linebreak \textit{Handbuch der Elektrizit\"at und des Mag\-ne\-tis\-mus: In f\"unf B\"anden / bearbeitet von\ldots Prof. Dr. M. v. Smoluchowski\ldots; hrsg. von Prof. Dr. L. Graetz} (Leipzig: Verlag von Johann Ambrosium Barth, \linebreak{} [1914]; 1921) {Band II: Station\"are Str\"ome}, S.~366--428.

\item \textrm{G\"ultigkeitsgrenzen des zweiten Hauptsatzes des W\"armetheorie [Limits of validity of the second law of thermodynamics]}. \textit{Vortr\"age \"uber die kinetische Theorie der Materie und der Elektrizit\"at, gehalten in G\"ottingen auf Einladung der Kommission der Wolfskehlstiftung} (Leipzig; Berlin: B. G. Teubner, 1914), S.~87--121.

\item[\textbf{1917}]

\item \textrm{Kobiety w naukach \'scis\l{}ych [Women in science; Lecture presented at the Union of Science and Literature in Lviv in 1912]}. \textit{Rok Polski}. 1917. \textbf{2}, S.~7--24.

\end{enumerate}

\medskip
Journal title abbreviations:

\begin{itemize}
{\it
\item[] Ann. d. Phys. =
Annalen der Physik, Vierte Folge

\item[] Bull. Int. Acad. Sci. Cracovie. Classe des sciences math. et naturelles =
Bulletin International de l'Aca\-d\'emie des Sciences de Cracovie. Classe des sciences math\'ematiques et naturelles. S\'erie A. Sciences math\'ematiques

\item[] Kosmos =
Kosmos: czasopismo polskiego Towarzystwa przyrodnik\'ow imienia Kopernika
[The journal of the Polish Copernicus Society of Natural Scientists]

\item[] Phil. Mag. Ser. 6 =
The London, Edinburgh, and Dublin Philosophical Magazine and Journal of Science, Sixth Series

\item[] Phys. Zs. =
Physikalische Zeitschrift

\item[] Rozprawy Wydzia\l{}u mat.-przyrod. Akad. Um. =
Rozprawy Wydzia\l{}u matematyczno-przy\-rod\-ni\-cze\-go Akademii Umiej\k{e}tno\'sci. Serya III
[Papers of the Section of mathematics and natural sciences of the  Academy of Sciences and Letters; {\rm this journal was issued as} Bulletin International\ldots\ {\rm (see above) with articles translated mostly into French and German and occasionally into English}\,]
}
\end{itemize}

\section{Not only a scientist}

Marian Smoluchowski was an active mountaineer and skier.
His first mountain experience was in 1884, at the age of twelve, when during summer vacations he climbed Obir (2100~m) in Carinthia~\cite{Palczewski97}. Alps became Smoluchowski's favorite place for mountain trips, where he returned many times.

\begin{figure}[h]
\centerline{%
\includegraphics[width=0.45\textwidth]{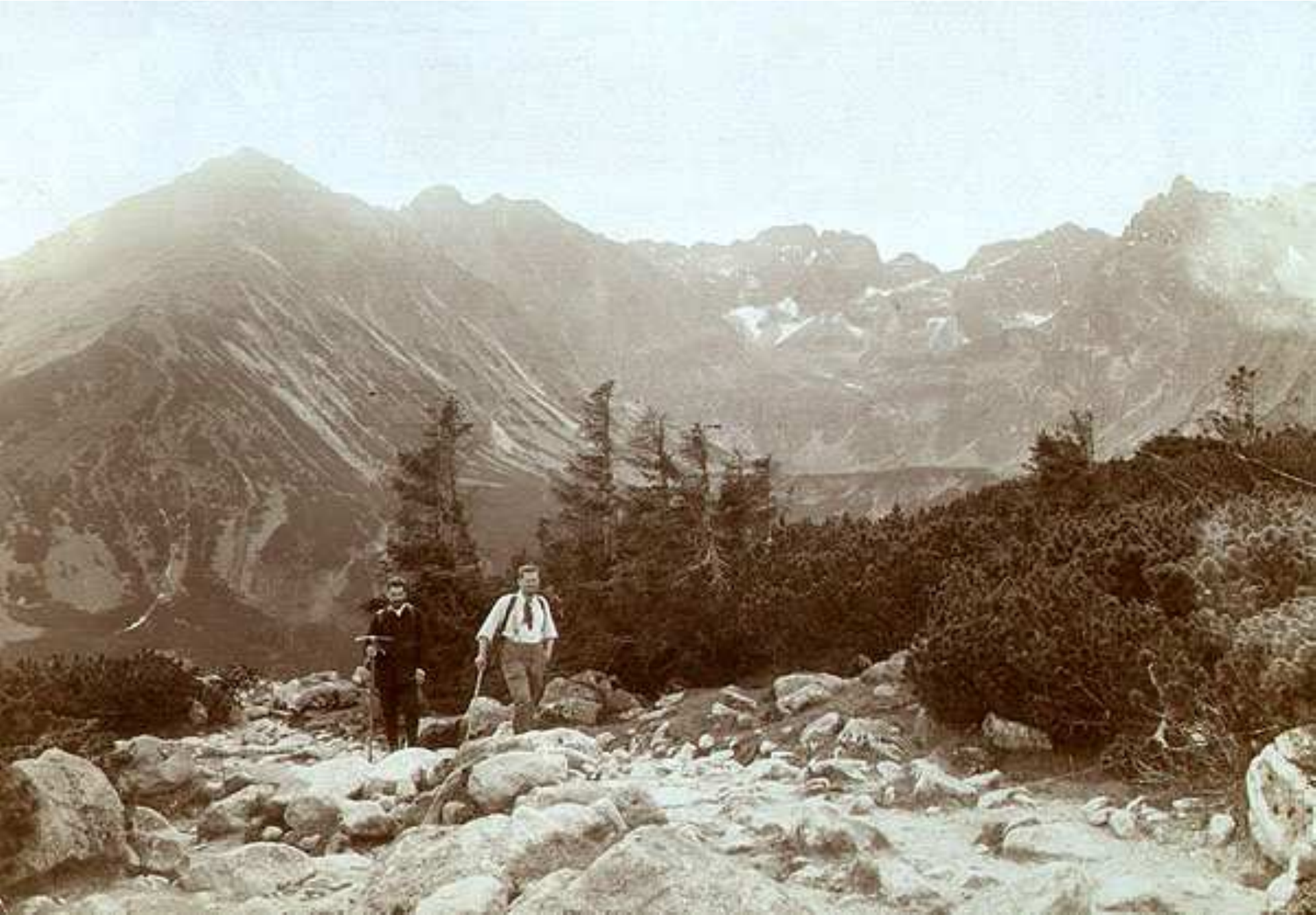}
\qquad
\includegraphics[width=0.45\textwidth]{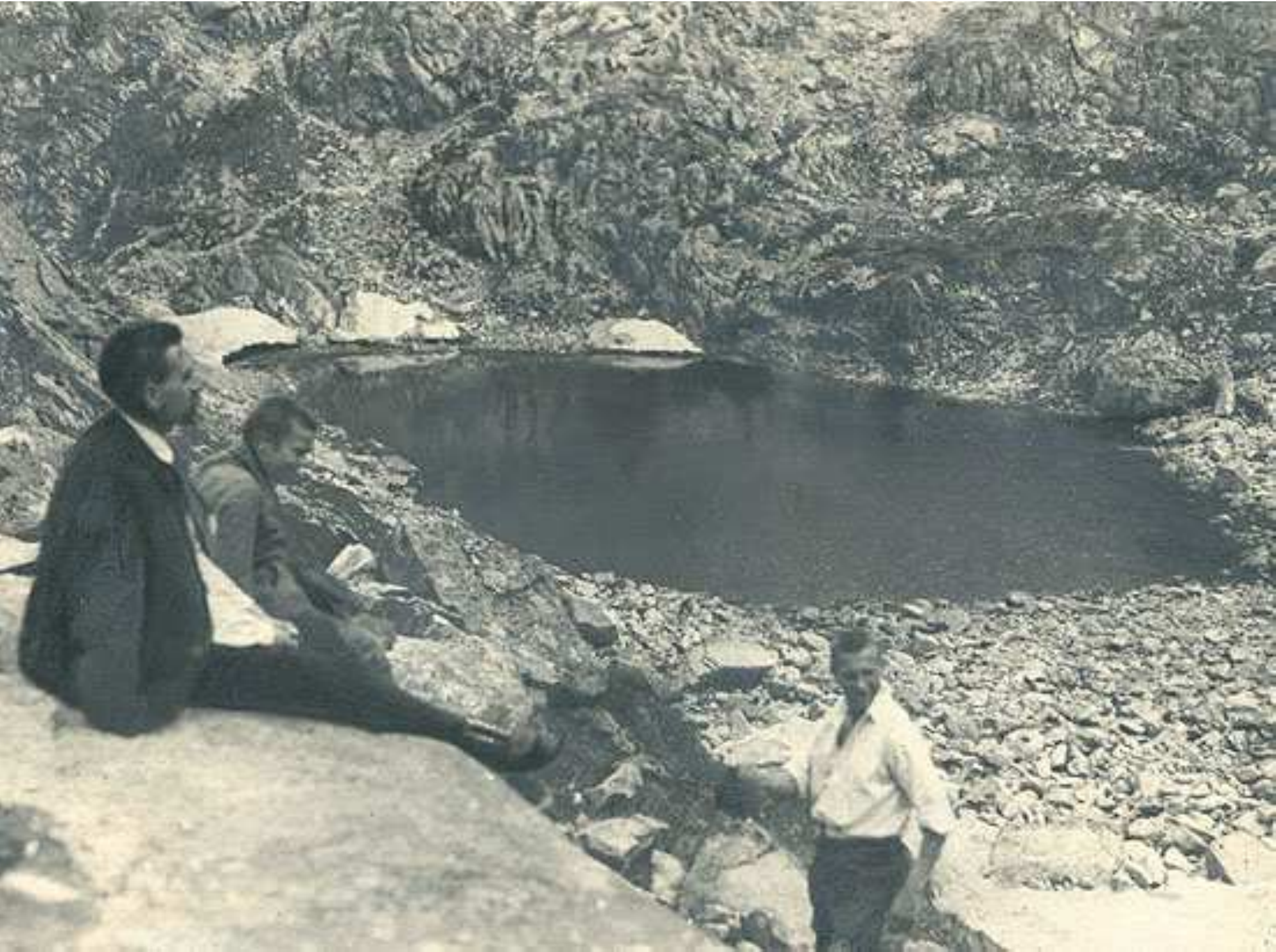}
}
\caption{Photos from Smoluchowski mountain trips~\cite{Grotowski03}.}
\label{fig:photos}
\end{figure}

In his numerous trips, both scientific and recreational, Smoluchowski visited a lot of places. Some of them were also mountainous: Pyrenees, Dolomites, Scottish Highlands, etc.~\cite{Goetel18}. Two photos from these trips are shown in figure~\ref{fig:photos}.

During his Lviv period, Smoluchowski mostly traveled in Tatra and Carpathian Mountains. He also started skiing in the mountains.
Many of his climbs were originally made during winter:
Syvulja (1836~m), Pikun (1657~m), Polensky (1693~m) in Gorgany;
Mih\u{a}ilecul (1926~m) i Farc\u{a}ul (1961~m) in Marmaros Carpathians.

In 1904 and 1909, Marian Smoluchowski together with his brother Tadeusz (1868--1936) traveled to Alps. In the second trip, also joined by Zygmunt Klemensiewicz and Tadeusz Kossowicz, they climbed Finsteraarhorn (4274~m), Jungfrau (4159~m), and Lauterbrunner Breithorn (3782~m)~\cite{Grotowski03}.

In 1908 Marian Smoluchowski became a member of the Tourist Section of the Polish Mountaineer Society, and in 1911/12 was the president of this Section \cite{Grotowski03,Michalski05}. In 1916 he was awarded the Silber Edelwei\ss\ by the German and Austrian Alpine Society (Deutsche und \"Osterreichische Alpenverein).

\begin{figure}[h]
\centerline{%
\includegraphics[width=0.45\textwidth]{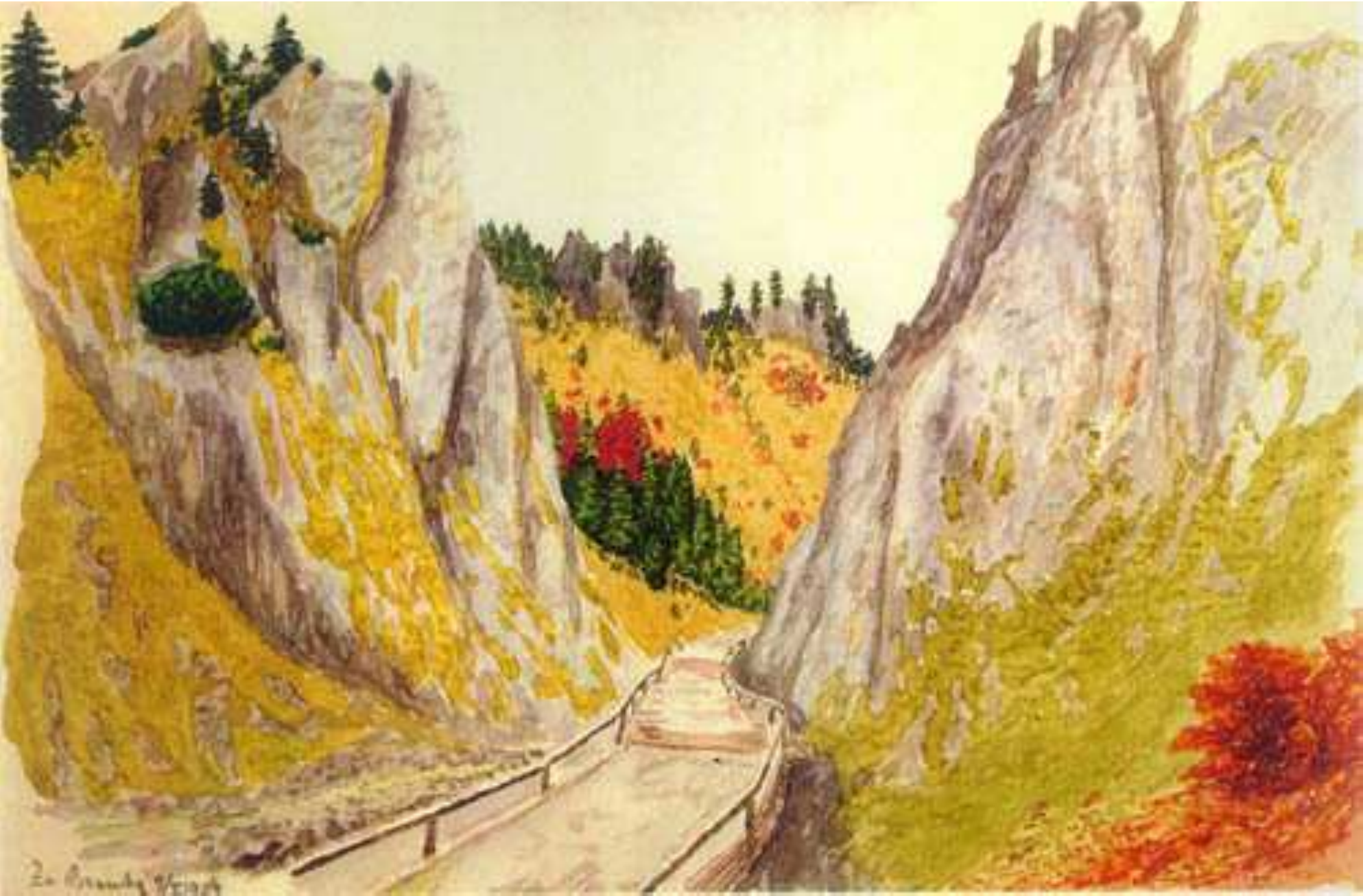}
\qquad
\includegraphics[width=0.45\textwidth]{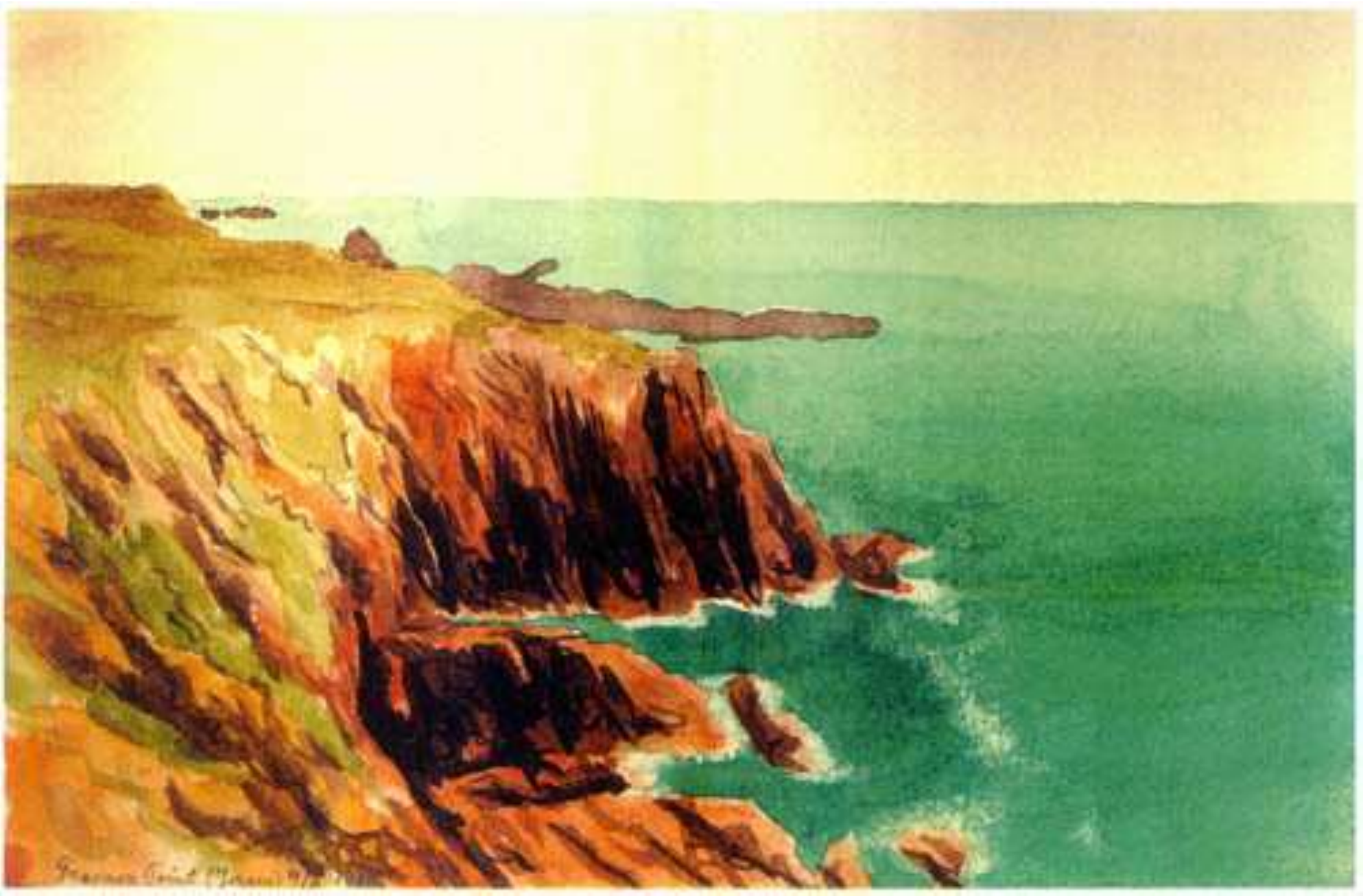}
}
\caption{Watercolor paintings by Marian Smoluchowski. Left:
\textit{Za bramk\k{a} [Beyond the gate]} (9 October 1904);
right:
\textit{Prosner Point, Jersey} (4 August 1908) \cite{Wyka03,Strzalkowski03}.
}
\label{fig:paintings}
\end{figure}

To complete the account on the personality of Marian Smoluchowski, one should mention his relation to art. He loved music and played the piano to relax, often together with Ms.~Jadwiga Baraniecka-Zakrzewska, his wife's sister~\cite[p.~254]{Teske55}.
Another Smoluchowski's hobby was painting. Some aquarelles drawn during his vacations are shown in figure~\ref{fig:paintings}.

\section*{Acknowledgements}
I would like to thank the staff of the University of Lviv Archive for making available the materials from the Archive stock. The staff of the Institute of Physics Library (Jagiellonian University) was very helpful in collecting the bibliography.


\begin{thebibliography}{99}
%
%
\bibitem{Teske55}
Teske~A., Marian Smoluchowski: \.zycie i tw\'orczo\'s\'c, PWN, Krak\'ow, 1955.

\bibitem{Teske77}
Teske~A., Marian Smoluchowski: Leben und Werk, Wroc\l{}aw--Warszawa--Krak\'ow--Gda\'nsk, 1977.

\bibitem{Teske68}
Teske~A., Szkic tw\'orczo\'sci Mariana Smoluchowskiego, In: Studia po\'swi\k{e}cone Marii Sk\l{}odowskiej-Curie i Marianowi Smoluchowskiemu, Wroc\l{}aw--Warszawa--Krak\'ow, 1968.

\bibitem{Sredniawa85}
\'Sredniawa~B., History of theoretical physics at Jagiellonian University in Cracow in XIXth century and in the first half of XXth century, Cracow, 1985.

\bibitem{Chandr00}Chandrasekhar~S., Kac~M., Smoluchowski~R., Marian Smoluchowski: his life and scientific work, PWN, Warsaw, 2000.


\bibitem{Pohl08}
Pohl~W.G.,
In: The Global and the Local: The History of Science and the Cultural Integration of Europe, Proceedings of the 2nd International Conference of the European Society for the History of Science (Cracow, 2006),  Kokowski~M. (Ed.), The Press of the Polish Academy of Arts and Sciences, Cracow, 2007, 419.


\bibitem{Hunger08}
Hunger~H.,
In: The Global and the Local: The History of Science and the Cultural Integration of Europe, Proceedings of the 2nd International Conference of the European Society for the History of Science (Cracow, 2006),  Kokowski~M. (Ed.), The Press of the Polish Academy of Arts and Sciences, Cracow, 2007, 412.

\bibitem{Rovenchak09}
Rovenchak~A.,
Acta Phys. Pol.~A, 2009, \textbf{116}, 109.

\bibitem{ArchiveLU}
Archive materials of the University of Lviv: F.~26, op.~5, spr.~1762.

\bibitem{Kronika12}Hahn~W.,
Kronika uniwersytetu lwowskiego. II. Lw\'ow, 1912.

\bibitem{AddressBook:1910}
Spiegel~J.R., Skorowidz adresowy krol. sto\l. miasta Lwowa, rok 1910.

\bibitem{AddressBook:1913}Reichman~F., Ksi\k{e}ga adresowa krol. sto\l. miasta Lwowa, rok 1913.

\bibitem{Map:1929}Plan miasta Lwowa, M\,1:15000,
Instytut Kartograficzny im.~E.~Romera, Lw\'ow--Warszawa, 1929.

\bibitem{Kronika99}
Kronika uniwersytetu lwowskiego, Lw\'ow, 1899.

\bibitem{Loria53}Loria~S.,
Post\k{e}py fizyki, 1953, \textbf{4}, 5.

\bibitem{Fulinski98}
Fuli\'nski~A.,
Acta Phys. Pol.~B, 1998, \textbf{29}, 1523.

\bibitem{new} \url{http://www.ktf.franko.lviv.ua/cgi-bin/select.cgi?Smoluchowski}.

\bibitem{Palczewski97}Palczewski~A.,
Delta: matematyczno-fizyczno-astronomiczny miesi\k{e}cznik, 1997, \textbf{12(283)}, 7.

\bibitem{Goetel18}Goetel~W.,
Pami\k{e}tnik Towarzystwa Tatrza\'nskiego, 1917/18, \textbf{36}, 42.

\bibitem{Grotowski03}
Grotowski~K.,
Zwoje: Periodyk kulturalny, 2003, \textbf{2(35)},
\url{http://www.zwoje-scrolls.com/zwoje35/text19p.htm}.

\bibitem{Michalski05}Michalski~Cz.,
Konspekt, 2005, \textbf{2(22)}, 183.

\bibitem{Wyka03}
Wyka~E.,
Zwoje: Periodyk kulturalny, 2003, \textbf{2(35)},
\url{http://www.zwoje-scrolls.com/zwoje35/text18p.htm}.

\bibitem{Strzalkowski03}
Strza\l{}kowski~A.,
Zwoje: Periodyk kulturalny, 2003, \textbf{2(35)},
\url{http://www.zwoje-scrolls.com/zwoje35/text16p.htm}.



\end{thebibliography}
%

 \ukrainianpart

\title{Львівський період Маріана Смолуховського}
\author{А. Ровенчак}
\address{Кафедра теоретичної фізики, Львівський національний університет імені Івана Франка, вул.~Драгоманова, 12, 79005 Львів, Україна}

\makeukrtitle

\begin{abstract}
\tolerance=3000%
Більшість наукових досягнень Маріана Смолуховського припадає на час його роботи у Львівському університеті. Оскільки ця сторона діяльності вченого добре описана в літературі, то у статті зроблено наголос на менш відомих напрямках роботи цього видатного вченого: викладання й орґанізаційні заходи і навіть його хобі. Наведено також перелік публікацій львівського періоду.

\keywords Маріан Смолуховський, Львівський університет

\end{abstract}

\end{document}